\documentclass[aps,pre,preprint,eqsecnum,amssymb,floatfix]{revtex4}
\usepackage{graphicx,color,psfrag}
\usepackage{amsmath}
\usepackage{bm}
\begin{document}
\title{Field-theoretical approach to open quantum systems and the Lindblad equation}
\author{Hans C. Fogedby}
\email{fogedby@phys.au.dk}
\affiliation{Department of Physics and
Astronomy, 
\\
University of Aarhus, Ny Munkegade,
\\
8000 Aarhus C, Denmark}
\begin{abstract}
We develop a systematic field-theoretical approach to open quantum systems based on condensed-matter
many-body methods. The time evolution of the reduced density matrix for the open quantum system 
is determined by a transmission matrix. Developing diagrammatic perturbation theory, invoking 
Wick's theorem in connection with a Caldeira-Leggett quantum oscillator environment
in thermal equilibrium, the transmission matrix satisfies a Dyson equation characterized by 
an irreducible kernel. 
Unlike the Nakajima-Zwanzig and standard approaches, the Dyson equation is equivalent to a general 
non-Markovian master equation for the reduced density matrix, incorporating secular effects
and independent of the initial preparation. The kernel is determined by a systematic
diagrammatic expansion in powers of the interaction. 
We consider the Born approximation for the kernel. 
Applying a condensed-matter pole or, equivalently, a quasiparticle-type approximation, equivalent to
the usual assumption of a timescale separation, we derive a master equation of the Markov type. 
Furthermore, imposing the rotating-wave approximation,we obtain a Markov master equation of 
the Lindblad form. To illustrate the method, we consider the standard example of a single 
qubit coupled to a thermal heat bath.
\end{abstract}
\maketitle
\section{\label{intro} Introduction}
There is a continuing strong interest in open quantum systems interacting with an environment  both from a
fundamental and an experimental point of view. The issue is important in many areas of science, spanning 
from chemistry over atomic, molecular, and optical physics to condensed-matter physics, quantum information,
and quantum computers \cite{Scully96,vanKampen92,Nielsen10,Feshbach58,Zanardi97,
Bourennane04,Verstraete09,Diehl11}. The theory of open systems has in particular been developed
in the field of quantum optics \cite{Breuer06,Walls94,Gardiner05}, where atoms or cavity modes 
are  coupled to the radiation field. In recent years there has been a strong focus on decoherence, entanglement, 
and dissipation in the context of nano quantum systems and quantum computation 
\cite{Aolita15,Bellomo07,Schlosshauer07,Schlosshauer19,Paneru20,Goold16,Davidovich16}.

There are a variety of theoretical approaches to open quantum systems
\cite{Breuer06,Gardiner05,Nakajima58,Wangsness53,Redfield65,Davies74,Majenz13,Mozgunov20,Rivas12,
Plenio98,Carmichael91,Walls94,Dalibard92,Moelmer93,Rammer91,Fogedby93}. 
Unlike the situation in condensed-matter physics where all degrees of freedom come into play
\cite{Mahan90,Bruus04}, open quantum systems have a finite number of degrees of
freedom, typically one or several qubits, atomic model systems, or cavity modes interacting with an 
environment or bath composed of many degrees of freedom. Like in classical statistical physics 
\cite{Landau80,Reichl98}, the bath is usually assumed to be unperturbed by the open quantum system; note, 
however, recent work on system-bath correlations \cite{Alipour20}. Defining the bath explicitly as a 
collection of independent quantum oscillators \cite{Caldeira83a,Caldeira83b} the properties of the 
bath are characterized by a spectral density and a temperature; the vacuum state corresponds 
to zero temperature.

The combined system, consisting of the open quantum system and the bath, constitutes a closed  quantum 
system evolving in time according to a unitary transformation \cite{Landau59,Breuer06}. On the other hand, 
the open quantum system evolves according to a non-unitary quantum map due to the interaction with 
the bath. This interaction gives rise to mixture, dissipation, decoherence, and entanglement
\cite{Breuer06}.

Since a mixture of states is generated by the system-bath coupling, the starting point for an analysis is 
usually based on the von Neumann equation $d\rho(t)/dt=-i[H,\rho(t)]$ \cite{Landau59,Breuer06} 
for the  density operator $\rho(t)$ for the combined closed system; here $H$ is the total Hamiltonian. 
Subsequently, the density operator for the open quantum system ({\it S}), $\rho_S(t)$, is obtained by tracing 
out the bath ({\it B}) degrees of freedom. Consequently, $\rho_S(t)=\text{Tr}_B[\rho(t)]$ and formally 
$d\rho_S(t)/dt=-i\text{Tr}_B[H,\rho(t)]$; note that throughout the present paper, we set $\hbar=1$.

In general the time evolution of the reduced density operator $\rho_S(t)$ is governed by an 
inhomogeneous integral equation, or, correspondingly, a master equation incorporating
memory effects. However, many studies of open quantum systems are based on the Markov assumption in
combination with the Born approximation. Similar to the Langevin - Fokker-Planck scheme in classical
statistical mechanics \cite{vanKampen92,Risken89,Reichl98}, the Markov approximation is based on 
an assumption of a separation of timescales. The intrinsic timescale of the bath, typically the fall-off time
of bath correlations, is assumed to be short in comparison with the timescale associated with the small 
open quantum system, characteristically given by the inverse level spacing. In this limit, we can ignore 
memory effects  and invoke the Markov approximation \cite{Breuer06,Gorini78}; note, however, recent work on 
non-Markovian issues \cite{Diosi14,Ferialdi16,Breuer16,Breuer09,Bonifacio20,Vega17,Ferialdi16}.  
Finally, in the second-order Born approximation it follows that $\rho_S(t)$ is governed by the Lindblad
equation \cite{Lindblad76,Manzano20},
\begin{eqnarray}
\frac{d}{dt}\rho_S(t)=-i[H_S,\rho_S(t)]+\sum_{\alpha\beta,k}\gamma_k^{\alpha\beta}
\Big[S^\beta_k\rho_S(t)S_k^{\alpha\dagger}-\frac{1}{2}\{S_k^{\alpha\dagger} S_k^\beta,\rho_S(t)\}\Big].
\label{Lindblad}
\end{eqnarray}
Here $H_S$ is the Hamiltonian for the open quantum system; note that here we ignore a renormalization
of $H_S$ due to a Lamb shift. The coupling to the bath is characterized by the jump operators $S_k^\alpha$
acting on the quantum system together with the positive dissipation coefficients $\gamma_k^{\alpha\beta}$. 
The commutator term characterizes the unitary time evolution in the absence of coupling to the bath according 
to the von Neumann equation $d\rho_S(t)/dt=-i[H_S,\rho_S(t)]$. The last term involving the 
jump operators $S_k^\alpha$ describes the coupling to the bath giving rise to dissipation, incoherence, 
and entanglement. The standard derivation of the Lindblad equation within the quantum optics approach 
is based on an iteration of the von Neumann equation to the second-order Born approximation 
in combination with a separation of timescales, the rotating-wave approximation,  and a time coarse graining.
The basic assumption is that bath correlations relax on a timescale much faster than the timescale associated 
with the evolution of the density operator; a standard source is the book 
by Breuer and Petruccione \cite{Breuer06}.

In the present paper, we approach the issue of open quantum systems and in particular the Lindblad equation
from the point of view of condensed-matter physics using techniques from equilibrium and non equilibrium 
many-body theory \cite{Abrikosov65,Kadanoff62,Mahan90,Bruus04,Keldysh65,Schwinger61}. 
We believe that the present many-body approach to an open quantum system, the Markov approximation, 
and in particular the Lindblad equation is sheds light on the assumptions underlying the standard 
approach \cite{Breuer06}. 

Below we briefly sketch our procedure. The time evolution of the reduced density operator $\rho_S(t)$ 
describing the open quantum system interacting with the bath can be expressed in the form 
$\rho_S(t)=T(t,t_i)\rho_S(t_i)$, where $T(t,t_i)$ is a transmission (super) operator transporting the density 
operator forward in time from an initial time $t_i$ to a final time $t$;  as commonly assumed,  the system is 
uncorrelated with the bath at the initial time $t_i$. Assuming that the bath is composed of non-interacting 
quantum oscillators and invoking Wick's theorem, it follows that the transmission operator $T(t,t')$ satisfies
a Dyson equation of the form $T(t,t')=T^0(t,t')+\int dt''dt'''T^0(t,t'')K(t'',t''')T(t''',t')$. Here $T^0(t,t')$ is 
the unperturbed transmission operator in the absence of coupling to the bath. The irreducible kernel or 
self-energy $K(t,t')$ can be determined perturbatively in terms of the coupling between the open quantum 
system and the bath. It then follows from the Dyson equation that the density operator satisfies the  master 
equation 
$d\rho_S(t)/dt = -i[H_S,\rho_S(t)]+\int dt' K(t,t')\rho_S(t')$. Here $H_S$ is the system Hamiltonian; 
we note that $T^0(t,t')$ solves  the von Neumann equation $d\rho_S(t)/dt=-i[H_S,\rho_S(t)]$. 
In general, the memory kernel $K(t,t')$ depends on the free evolution of the system, $\exp(-iH_St)$, and 
correlations characterizing the bath. In the Born approximation to second order in the system-bath 
coupling, the kernel $K(t,t')$ consists of four terms including a single bath correlation function.  
To implement the Markov approximation, we note that the behavior of $\rho_S(t)$ is determined by the 
transmission operator $T(t,t')$ or in Fourier space the pole structure of $\tilde T(\omega)$. From the 
Dyson equation for $\tilde T(\omega)$ we obtain formally $\tilde T(\omega)^{-1}=
[\tilde T^0(\omega)^{-1}-\tilde K(\omega)]$ and the resonance structure is determined by 
$\text{det}[\tilde T^0(\omega)^{-1}-\tilde K(\omega)]=0$. Implementing from many-body theory the 
so-called quasiparticle approximation to leading order in the system-bath coupling by setting 
$\text{det}[\tilde T^0(\omega)^{-1}]=0$,  we obtain a set of frequencies  $\omega_n$. In this 
approximation setting $\tilde K(\omega)=\tilde K(\omega_n)$ we obtain a Markov equation. 
Imposing furthermore the rotating-wave approximation, we arrive at the Lindblad equation. 

Aiming at a self-contained exposition, the paper is organized as follows. In Sec. \ref{general} we set up
the general scheme deriving an expression for the reduced density matrix in terms of a transmission matrix; 
details are deferred to  Appendix \ref{app1}. In Sec. \ref{bath} we introduce the Caldeira-Leggett heat bath 
in terms of independent quantum oscillators, and we derive the bath correlation functions. Moreover, we introduce
Wick's theorem with a derivation deferred to Appendix \ref{app2}. Sec. \ref{transmission} is devoted to
a discussion of the transmission matrix and a derivation of the Dyson equation. In  Sec. \ref{master} 
we present the general master equation for the reduced density matrix following from the Dyson
equation. In  \ref{born} we derive the general non-markovian master equation to second order in the coupling
between the system and the bath, i.e., the Born approximation. In Sec. \ref{lindblad} the Lindblad equation
is reviewed . In Sec. \ref{standard} we summarize the standard microscopic derivation of the
Lindblad equation. In Sec. \ref{qp1} we present a heuristic derivation of 
the Lindblad equation, in Sec. \ref{qp2} a discussion of the pole structure of the transmission matrix, and in  
Sec. \ref{qp3}  
a detailed discussion of the quasiparticle 
approximation, resulting in a Markovian master equation. In Sec. \ref{dis1} we present 
a discussion of 
the equivalence between the field-theoretical approach and the standard derivation. Moreover, we discuss
aspects of the timescale separation: in Sec. \ref{dis2} we discuss the approach by Diosi and Ferialdi 
\cite{Diosi14,Ferialdi16} and its relation to the present work, in  Sec. \ref{dis3}
the Nakajima - Zwanzig approach to a non-Markovian master equation, and finally in 
Sec. \ref{dis4} a discussion of the quasiparticle approximation in the simple case of a  
qubit coupled to a bath. In Sec. \ref{summary} we give a short summary of our results.
In Appendix \ref{app1} we discuss the interaction representation, and in Appendix \ref{app2} we present a 
derivation of Wick's theorem.
\section{\label{general} General analysis}
Here we set up the field-theoretical approach to open quantum systems. Details regarding the interaction
representation are deferred to Appendix \ref{app1}. The methods used are well-known in condensed-matter theory 
both for equilibrium \cite{Abrikosov65,Mahan90,Bruus04} and non equilibrium systems \cite{Keldysh65,
Kadanoff62,Schwinger61}, but they have not been used much in the context of open quantum systems
\cite{Breuer06,Gardiner05,Walls94}; see, however, recent work in \cite{Ferguson21,Sieberer16}. 
In condensed-matter theory, the essential building block is the Green's function or propagator in the 
context of diagrammatic or functional perturbative expansions. These schemes have been developed 
both for equilibrium and nonequilibrium systems.

For open quantum systems, the standard tool is the reduced density operator $\rho_S(t)$ obtained by tracing
out the degrees of freedom of the environment. Expectation values of observables $A_S$ relating to
the open quantum system are thus given by $\langle A_S\rangle(t)=\text{Tr}_S[ \rho_S(t)A_S]$, which together
with the regression theorem provide the standard tools in quantum optics \cite{Breuer06,Gardiner05} and related
fields. In discussing open quantum systems, the starting point is a small system ({\it S}) with a finite number of 
degrees of freedom, typically one or several qubits, an atomic system or a cavity mode characterized by a 
quantum oscillator, interacting with a large quantum system acting as a bath ({\it B}) or an environment composed 
of many or infinitely many degrees of freedom. The total system composed of the open quantum system 
interacting with the quantum bath is thus described by the Hamiltonian
\begin{eqnarray}
&&H=H_S+H_B+H_{SB},
\\
&&H_{SB}=\sum_\alpha S^{\alpha}B^{\alpha}=\bm{S}\cdot\bm{B}.
\label{Hamiltonian} 
\end{eqnarray}
Here $H_S$ is the Hamiltonian for the quantum system,  $H_B$ the Hamiltonian for the bath, and $H_{SB}$ 
the interaction describing the coupling between the quantum system and the bath. The system and the bath 
live in separate Hilbert spaces, i.e., $[H_B,H_S]=0$; however, due to the interaction we have
$[H_S,H_{SB}]\neq 0$ and $[H_B,H_{SB}]\neq 0$. $S^\alpha$ are operators acting on the quantum system 
and $B^\alpha$ operators monitoring the bath; $\bm{S}\cdot\bm{B}$ denotes the usual scalar product.
Since $H_{SB}^\dagger=H_{SB}$ we have 
$\sum_\alpha S^{\alpha}B^{\alpha}=\sum_\alpha S^{\alpha\dagger}B^{\alpha\dagger}$; note that this
relation does not imply that $S^\alpha$ or $B^\alpha$ individually are Hermitian.

The coupling between the quantum system and the bath gives rise to mixed states, relaxation, decoherence 
and entanglement between system and bath and it is customary to use the density operator formalism. 
The density operator has the form $\rho(t)=\sum_nP_n|\Psi_n(t)\rangle\langle\Psi_n(t)|$, where the 
probabilities $P_n\ge 0$ and $\sum_n P_n=1$ \cite{vonNeumann27}. The density operator is Hermitian, 
positive, and has unit trace, i.e., $\rho(t)^\dagger=\rho(t)$, $\langle\Phi|\rho(t)|\Phi\rangle>0$, and 
$\text{Tr}[\rho(t)]=1$. The purity of a quantum state is defined as $\gamma(t)=\text{Tr}[\rho(t)^2]$; for a 
pure state $\gamma=1$ for a mixed state $\gamma<1$. 

It follows from the Schr\"{o}dinger equation, 
$id|\Psi(t)\rangle/dt=H|\Psi(t)\rangle$ \cite{Landau59}, that the density operator satisfies 
the von Neumann equation  \cite{vonNeumann27,Landau27}
\begin{eqnarray}
\frac{d}{dt}\rho(t)=-i[H,\rho(t)],
\label{vonNeumann}
\end{eqnarray}
with formal solution
\begin{eqnarray}
&&\rho(t)=U(t,t_i)\rho(t_i)U(t,t_i)^\dagger,
\label{Den-Sol}
\\
&&U(t,t')=\exp(-iH(t-t')).
\label{Uni-Evol}
\end{eqnarray}
Here $\rho(t_i)$ is the density operator at an initial time $t_i$ and $U(t,t')$ is the usual unitary evolution 
operator. From (\ref{Den-Sol}) we infer the density operator 
for the system $\rho_S(t)$ by tracing over the bath states, i.e.,
\begin{eqnarray}
\rho_S(t)=\text{Tr}_B[\rho(t)].
\label{Density-operator}
\end{eqnarray}
In recent work \cite{Alipour20} the issue of correlations between system
and bath has been addressed. However, in the present context we assume that the bath and system are
uncorrelated at an initial time $t_i$ and that the bath composed of many degrees of freedom is maintained
in a stationary thermodynamic state characterized by the density operator 
$\rho_B=\exp(-\beta H_B)/\text{Tr}[\exp(-\beta H_B)]$
\cite{Reichl98}; here $\beta$ is the inverse temperature; the vacuum state corresponding to $\beta=\infty$. 
The initial condition is thus given by the factorised density operator
\begin{eqnarray}
\rho(t_i)=\rho_B\rho_S(t_i).
\label{Initial-Condition}
\end{eqnarray}
In terms of a super or Liouville operator $L$ the time evolution in (\ref{Den-Sol}) can be written in the form
$\rho(t)=L(t,t_i)\rho(t_i)$, where in a complete basis $|n\rangle$ we have 
$\rho(t)_{pp'}=\sum_{qq'}L_{pp',qq'}(t,t_i)\rho(t_i)_{qq'}$ and $L_{pp',qq'}(t,t_i)=
U(t,t_i)_{pq}U(t,t_i)^\dagger_{q'p'}$. Likewise, due to the linearity the time evolution of 
the reduced density operator $\rho_S(t)=\text{Tr}_B[L(t,t_i)\rho_B\rho_S(t_i)]$ 
for the system can be expressed in a similar form 
\begin{eqnarray}
&&\rho_S(t)=T(t,t_i)\rho_S(t_i),
\label{Trans-Op}
\\
&&\rho_S(t)_{pp'}=\sum_{qq'}T(t,t_i)_{pp',qq'}\rho_S(t_i)_{qq'},
\label{Trans-Mat}
\end{eqnarray}
where  $T(t,t_i)$ is a transmission (super) operator determining the time evolution of $\rho_S(t)$.

To proceed systematically, we turn to an expansion of the evolution operator $U(t,t_i)$
in powers of the system-bath interaction $H_{SB}$. Introducing the interaction representation
\cite{Bruus04,Mahan90} with respect to the unperturbed system and bath we set
\begin{eqnarray}
H_0=H_S+H_B,
\label{Interaction_Rep}
\end{eqnarray}
and we obtain in the interaction picture
\begin{eqnarray}
&&H_{SB}(t)=\exp(iH_0t)H_{SB}\exp(-iH_0t),
\label{Inter- HSB}
\\
&&\bm{B}(t)=\exp(iH_Bt)\bm{B}\exp(-iH_Bt),
\label{Inter-B}
\\
&&\bm{S}(t)=\exp(iH_St)\bm{S}\exp(-iH_St).
\label{Inter-S}
\end{eqnarray}
Expanding $U(t,t')$ and the Hermitian conjugate  $U(t,t')^\dagger$ in powers of  $H_{SB}(t)$ we have 
\begin{eqnarray}
&&U(t,t')=
\nonumber
\\
&&+i\sum_{n=0}\int dt_ndt_{n-1}\cdots dt_1G_R(t,t_n)
\bm{S}_{n}G_R(t_{n},t_{n-1})\bm{S}_{n-1}\cdots\bm{S}_{2}G_R(t_{2},t_{1})\bm{S}_{1}G_R(t_1,t')\times 
\nonumber
\\
&&e^{-iH_Bt}\bm{B}_{n}(t_n)\bm{B}_{n-1}(t_{n-1})\cdots\bm{B}_{2}(t_{2})\bm{B}_{1}(t_{1})e^{iH_Bt'},
\label{U-Exp}
\\
&&U(t,t')^\dagger=
\nonumber
\\
&&-i\sum_{n=0}\int du_ndu_{n-1}\cdots du_1G_A(t',u_1)
\bm{S}_{1} G_A(u_{1},u_{2})\bm{S}_{2}\cdots\bm{S}_{n-1}G_A(u_{n-1},u_{n})\bm{S}_{n}G_A(u_n,t)\times 
\nonumber
\\
&&e^{-iH_Bt'}\bm{B}_{1}(u_1)\bm{B}_{2}(u_{2})\cdots\bm{B}_{n-1}(u_{n-1})\bm{B}_{n}(u_{n})e^{iH_Bt},
\label{U-Dag-Exp}
\end{eqnarray}
where we have used $\bm{S}\cdot\bm{B}=(\bm{S}\cdot\bm{B})^\dagger$, introduced 
$H_{SB}(t_p)=\exp(iH_St_p){\bm S_p}\exp(-iH_St_p){\bm B_p}(t_p)$, and the retarded and 
advanced system Green's functions
\begin{eqnarray}
&&G_R(t,t')=-i\eta(t-t')\exp(-iH_S(t-t')),
\label{Retard-Green}
\\
&&G_A(t,t')=+i\eta(t'-t)\exp(-iH_S(t-t'));
\label{Advan-Green}
\end{eqnarray}
here the step function is given by $\eta(t)=1$ for $t>0$ and $\eta(t)=0$ for $t<0$, $\eta(0)=1/2$.
We also note that the step functions ensure that the integration range is from $t'$ to $t$ thus ensuring
causality, i.e., choosing the solution progressing forward in time.
By insertion in (\ref{Den-Sol}) the global density operator assumes the form
\begin{eqnarray}
\rho(t)=
\sum_{n=0,m=0}^\infty
&&\int dt_n\cdots dt_1G_R(t,t_n)\bm{S}_nG_R(t_n,t_{n-1})\cdots G_R(t_2,t_1)\bm{S}_1G_R(t_1,t_i)
\rho_S(t_i)
\times
\nonumber
\\
&&\int du_1\cdots du_mG_A(t_i,u_1)\bm{S}_1G_A(u_1,u_{2})\cdots G_A(u_{m-1},u_m)\bm{S}_mG_A(u_m,t)
\times
\nonumber
\\
&&e^{-iH_Bt}\bm{B}_{n}(t_n)\cdots\bm{B}_{1}(t_{1})\rho_B
\bm{B}_{1}(u_1)\cdots\bm{B}_{m}(u_{m})e^{iH_Bt};
\label{Global-Den}
\end{eqnarray}
note that this expression is completely equivalent to (\ref{Den-Sol}). For the reduced density operator 
$\rho_S(t)$ tracing over the bath and permuting the bath operators we obtain
\begin{eqnarray}
\rho_S(t)=
\sum_{n=0,m=0}^\infty
&&\int dt_n\cdots dt_1G_R(t,t_n)\bm{S}_n\cdots\bm{S}_1G_R(t_1,t_i)
\rho_S(t_i)\times
\nonumber
\\
&&\int du_1\cdots du_mG_A(t_i,u_1)\bm{S}_1\cdots\bm{S}_mG_A(u_m,t)\times
\nonumber
\\
&&\text{Tr}_B[\rho_B\bm{B}_1(u_1)\cdots\bm{B}_m(u_m)\bm{B}_n(t_n)\cdots\bm{B}_1(t_1)].
\label{Red-Den}
\end{eqnarray}
We note that the retarded and advanced Green's functions ensure that the operators $\bm S_p$ and 
$\bm B_p$  in 
(\ref{Red-Den}) are chronologically ordered along the so-called Keldysh contour from 
time $t$ to the initial time $t_i$ and back to time $t$ \cite{Keldysh65,Schwinger61,Rammer86}. 
We also remark that the expansion (\ref{Red-Den}) is completely equivalent to 
$\rho_S(t)=\text{Tr}_B[U(t,t_i)\rho(t_i)U(t_i,t)]$. Finally, inserting in (\ref{Trans-Mat}) we obtain for the
transmission matrix
\begin{eqnarray}
T(t,t')_{pp',qq'}=\sum_{n=0,m=0} &&\int dt_n\cdots dt_1(G_R(t,t_n)\bm{S}_n
\cdots\bm{S}_1G_R(t_1,t'))_{pq}\times
\nonumber
\\
&&\int du_1\cdots du_m(G_A(t',u_1)\bm{S}_1\cdots\bm{S}_mG_A(u_m,t))_{q'p'}\times
\nonumber
\\
&&\text{Tr}_B[\rho_B\bm{B}_1(u_1)\cdots\bm{B}_m(u_m)\bm{B}_n(t_n)\cdots\bm{B}_1(t_1)].
\label{Trans-Exp}
\end{eqnarray}
In the unperturbed case, we have in particular
\begin{eqnarray}
T^0(t,t')_{pp',qq'}=G_R(t,t')_{pq}G_A(t',t)_{q'p'},
\label{Trans-Unper}
\end{eqnarray}
describing the unperturbed propagation of $\rho_S(t)_{pp'}$.
Note that $\bm{B}_p(t_p)$ forms a scalar product with  $\bm{S}_p$ positioned between $G_R(t_{p+1},t_{p})$ 
and $G_R(t_{p},t_{p-1})$; likewise, $\bm{B}_q(u_q)$ is associated with  $\bm{S}_q$ positioned between 
$G_A(u_{q-1},u_{q})$ and $G_A(u_{q},u_{q+1})$.

The expression (\ref{Trans-Exp}) provides a direct expansion of the transmission matrix $T(t,t')$  in 
powers of the interaction $H_{SB}$  in terms of the Green's functions for the quantum system 
and the multitime bath correlations $\text{Tr}_B[\rho_B\bm{B}_1(t_1)\bm{B}_2(t_2)\bm{B}_3(t_3)\cdots]$. 
We already here note that since $\rho_S(t)=T(t,t_i)\rho_S(t_i)$ we have $d\rho_S(t)/dt=(dT(t,t_i)/dt)\rho_S(t_i)$ 
and it follows that the expansion of $T$ does not yield a proper master equation, independent of the initial 
condition $\rho_S(t_i)$. As a matter of fact, the expansion does not account for secular effects unless we 
make further
assumptions regarding the heat bath characterized by the density operator $\rho_B$ and the operators $\bm{B}$ 
monitoring the bath.
\section{\label{bath} Bath}
At this stage the expression (\ref{Trans-Exp}) for the transmission matrix applies to a general bath characterized 
by the multi-bath correlations $\text{Tr}_B[\rho_B{\bm B_1}(t_1)\cdots {\bm B_m}(t_m)]$. In order to proceed and 
account for secular effects and in this connection invoke Wick's theorem, we specify the structure of the bath 
according to the Caldeira-Leggett prescription \cite{Caldeira83a,Caldeira83b}. Details regarding Wick's theorem 
are deferred to Appendix \ref{app2}.
\subsection{\label{bath-correlations}Bath correlations}
For simplicity we consider a single bath with a bosonic structure, i.e., a collection of independent 
quantum oscillators, characterized by the bath Hamiltonian
\begin{eqnarray}
H_B=\sum_k\Omega_k n_k,
\label{Bath-Hamiltonian}
\end{eqnarray}
where the occupation number $n_k=b_k^\dagger b_k$ and $b_k$ is a Bose field with commutator 
$[b_k,b_p^\dagger]=\delta_{kp}$. The frequency associated with the wavenumber $k$ is $\Omega_k$. 
We assume that the bath is maintained at a temperature $1/\beta$ and characterized by the density operator 
\cite{Reichl98}
\begin{eqnarray}
\rho_B=\frac{\exp(-\beta H_B)}{\text{Tr}_B[-\exp(\beta H_B)]}.
\label{Den-Bath}
\end{eqnarray}
From the Heisenberg equation of motion, $idb_k/dt=[b_b,H_B]$ \cite{Bruus04,Mahan90} we infer
\begin{eqnarray}
&&b_k(t)=b_k\exp(-i\Omega_k t),
\label{b-Evol}
\\
&&b_k^\dagger(t)=b_k^\dagger\exp(+i\Omega_k t).
\label{b-dag-Evol}
\end{eqnarray}
Moreover, the mean occupation number or Planck distribution \cite{Reichl98,Landau80} is given by
\begin{eqnarray}
\langle n_k\rangle=\text{Tr}_B[\rho_B n_k]=\frac{1}{\exp(\beta\Omega_k)-1}.
\label{Planck}
\end{eqnarray}
For the bath operators $B^\alpha$ entering in the coupling in (\ref{Hamiltonian}) we choose for a single reservoir
$\bm{B}=(B^1,B^2)$, where
\begin{eqnarray}
&&B^1(t)=B(t)=\sum_k\lambda_kb_k(t),
\label{B1}
\\
&&B^2(t)=B^\dagger(t)=\sum_k\lambda_kb_k^\dagger(t);
\label{B2}
\end{eqnarray}
note that the coupling constants $\lambda_k$ can be chosen real by an appropriate choice of 
the phases of $b_k$. With this assignment we have for the relevant bath correlations
\begin{eqnarray}
&&D^{12}(t,t')=\text{Tr}_B[\rho_B B(t)B^\dagger(t')]=
\sum_k\lambda_k^2(1+\langle n_k\rangle)\exp(-i\Omega_k(t-t')),
\label{Bath-D12	}
\\
&&D^{21}(t,t')=\text{Tr}_B[\rho_B B^\dagger(t)B(t')]=\sum_k\lambda_k^2\langle n_k\rangle\exp(+i\Omega_k(t-t')).
\label{Bath-D21}
\end{eqnarray}
Introducing the spectral density of states $g(\Omega)$ and the Planck distribution $n(\Omega)$
\begin{eqnarray}
&&g(\Omega)=2\pi\sum_k\lambda_k^2\delta(\Omega-\Omega_k),~~\Omega>0,
\label{Spectral-Dis}
\\
&&n(\Omega)=\frac{1}{\exp(\beta\Omega)-1}, ~~\Omega>0,
\label{Planck2}
\end{eqnarray}
we also have
\begin{eqnarray}
&&D^{12}(t,t')=\int_0\frac{d\Omega}{2\pi} g(\Omega)(1+n(\Omega))\exp(-i\Omega(t-t')),
\label{Bath1-Corr}
\\
&&D^{21}(t,t')=\int_0\frac{d\Omega}{2\pi}g(\Omega)n(\Omega)\exp(+i\Omega(t-t')),
\label{Bath2-Corr}
\end{eqnarray}
and introducing the Fourier transform, $D^{\alpha\beta}(t,t')=
\int(d\omega/2\pi)\exp(-i\omega(t-t'))\tilde D^{\alpha\beta}(\omega)$,
\begin{eqnarray}
&&\tilde D^{12}(\omega)=g(\omega)(1+n(\omega)), ~~~\omega>0,
\label{Bath1-Fourier}
\\
&&\tilde D^{21}(\omega)=g(-\omega)n(-\omega),~~~ \omega<0.
\label{Bath2-Fourier}
\end{eqnarray}
Note that the system-bath coupling $\lambda_k$ is incorporated in the definition of the spectral density 
$g(\omega)$ which is of second order in $\lambda_k$. We also observe that in the classical limit for 
$\beta\to\infty$ we have $n(\omega)\to 1/\beta\Omega$ and the ohmic approximation requires
$g(\Omega)\propto\Omega$ (with an appropriate high frequency cut-off).
%
\subsection{\label{wick}Wick's theorem}
For a bath composed of independent quantum oscillators or bosons and assuming that the bath operator $\bm B$
is a  a linear combination of Bose creation and annihilation operators and, moreover, assuming that the density 
operator $\rho_B$ either corresponds to the Bose vacuum, i.e., $\rho_B=|0\rangle\langle 0|$ or a bath 
described by the density operator $\rho_B=\exp(-\beta H_B)/\text{Tr}_B[\exp(-\beta H_B)]$, Wick's theorem
holds for the multi-bath correlations in (\ref{Trans-Exp}).

Wick's theorem \cite{Gaudin60,Bruus04,Mahan90,Zinn-Justin89,Rammer91} implies that the multi-bath 
correlations $\text{Tr}_B[\rho_B{\bm B_1}(t_1)\cdots {\bm B_m}(t_m)]$ can be broken up into all possible 
pairings or contractions; we note that $\text{Tr}_B[\rho_B{\bf B(t)}] =0$.  In the case of four bath operators 
we have for example
\begin{eqnarray}
&&\text{Tr}_B[\rho_B B^\alpha(t_1)B^\beta(t_2)B^\gamma(t_3)B^\delta(t_4)]=
\nonumber
\\
&&D^{\alpha\beta}(t_1,t_2)D^{\gamma\delta}(t_3,t_4)+D^{\alpha\gamma}(t_1,t_3)D^{\beta\delta}(t_2,t_4)+
D^{\alpha\delta}(t_1,t_4)D^{\beta\gamma}(t_2,t_3),
\label{Wick-Four}
\end{eqnarray}
where we note that the time ordering is preserved; we have introduced the bath correlation function
\begin{eqnarray}
D^{\alpha\beta}(t,t')=\text{Tr}_B[\rho_BB^\alpha(t)B^\beta(t')].
\label{Bath-Corr}
\end{eqnarray}
Wick's theorem is basically a generalisation of the contraction of Gaussian multi-correlation
functions in terms of a single correlation function \cite{Zinn-Justin89} to noncommuting operators.
Usually the derivation of Wick's theorem is applied to the  vacuum expectation value of time-ordered
products or in the finite temperature case imaginary time-ordered products \cite{Bruus04,Mahan90}.
In the present context we summarize in Appendix \ref{app2} an elegant proof by Gaudin \cite{Gaudin60}
directly applied to a thermal average of operator products relevant to the present analysis. We also state
Wick's theorem in generator form \cite{Zinn-Justin89}.
\section{\label{transmission} Transmission matrix}
\begin{figure}[p]
\begin{center}
\includegraphics[width=0.9\hsize]{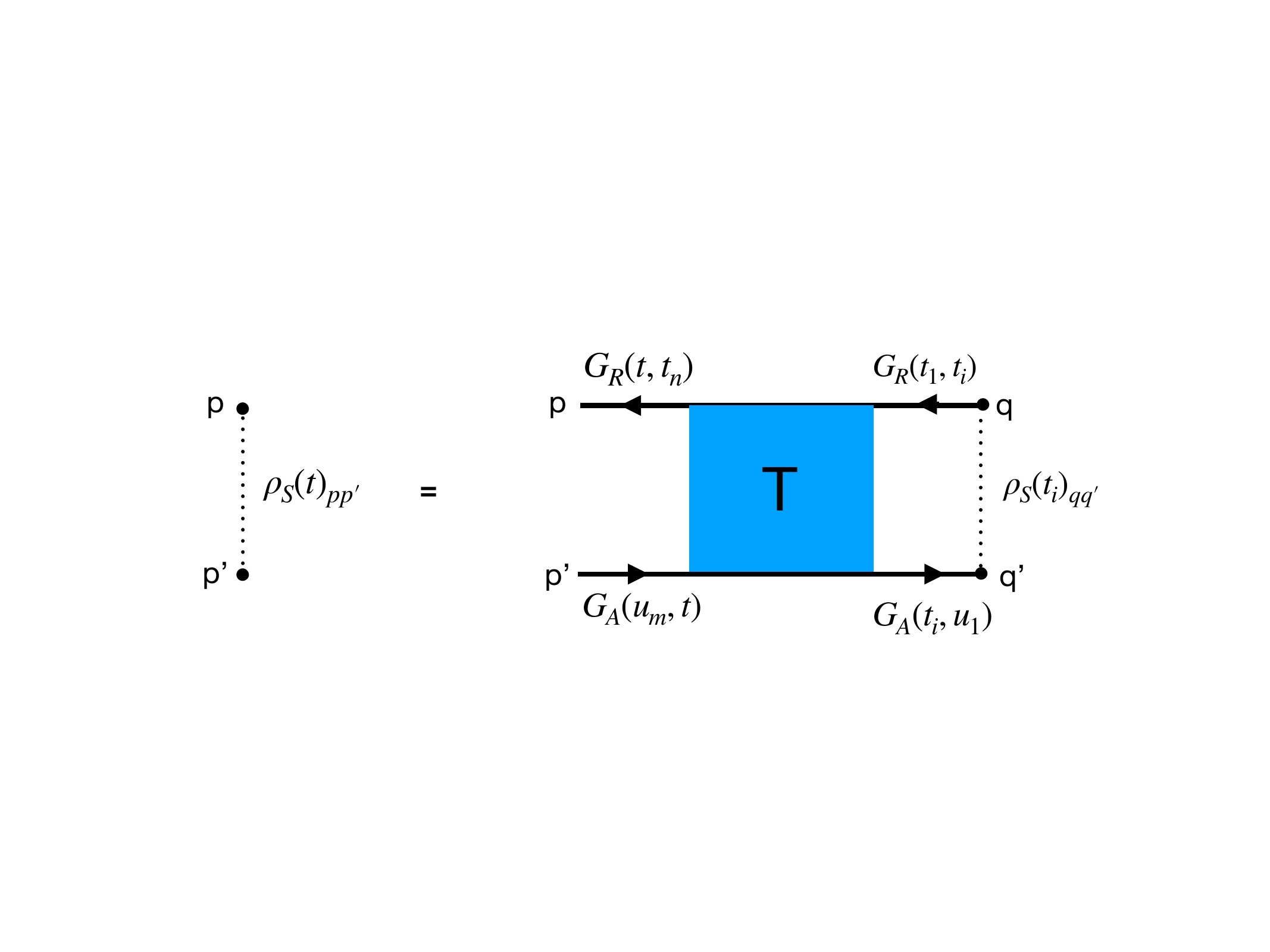}
\end{center}
\caption{Here we depict the transmission matrix $T(t,t_i)$ shown as a shaded box describing the evolution of
the reduced density operator $\rho_S(t)$ from the initial time $t_i$ to the final time $t$. The legs
on the reducible kernel $M$, and the retarded and advanced Green's functions $G_R$ and $G_A$, are denoted
by directed arrows.}
\label{fig1}
\end{figure}

The transmission operator $T(t,t')$ is of central importance in our analysis; 
With an explicit bath prescription and Wick's theorem, we  are in a position to discuss the transmission 
matrix (\ref{Trans-Exp}) in more detail. Inserting the identity $\eta(t)+\eta(-t)=1$, we can express (\ref{Trans-Exp}) 
in the form
\begin{eqnarray}
T(t,t')_{pp',qq'}=\sum_{n=0,m=0,ll',nn'} &&\int dt_n\cdots dt_1G_R(t,t_n)_{pl}(\bm{S}_n
\cdots\bm{S}_1)_{ll'}G_R(t_1,t')_{l'q}\times
\nonumber
\\
&&\int du_1\cdots du_mG_A(t',u_1)_{q'n}(\bm{S}_1\cdots\bm{S}_m)_{nn'}G_A(u_m,t)_{n'p'}\times
\nonumber
\\
&&(\eta(t_n-u_m)+\eta(u_m-t_n))(\eta(t_1-u_1)+\eta(u_1-t_1))\times
\nonumber
\\
&&\text{Tr}_B[\rho_B\bm{B}_1(u_1)\cdots\bm{B}_m(u_m)\bm{B}_n(t_n)\cdots\bm{B}_1(t_1)],
\label{Trans-Exp2}
\end{eqnarray}
yielding the sum of four individual contributions. Considering, for example the term containing the product  
$\eta(t_n-u_m)\eta(t_1-u_1)$ and using the identities
\begin{eqnarray}
G_R(t,t')_{pq}=+i\sum_lG_R(t,t'')_{pl}G_R(t'',t')_{lq},
\label{GR-Iden}
\\
G_A(t,t')_{pq}=-i\sum_lG_A(t,t'')_{pl}G_A(t'',t')_{lq},
\label{GA-Iden}
\end{eqnarray}
following from the definitions (\ref{Retard-Green}) and (\ref{Advan-Green}),
we make the replacements 
\begin{eqnarray}
&&G_R(t_1,t')=iG_R(t_1,u_1)G_R(u_1,t'),
\\
&&G_A(u_m,t)=-iG_A(u_m,t_n)G_A(t_n,t),
\label{replacements}
\end{eqnarray}
and we include the dummy arguments $u_1$ and $t_n$ in the integrations over $u_1$ and $t_n$. 
Applying this procedure to all contributions, we can express the transmission matrix in the form
\begin{eqnarray}
T(t,t')_{pp',qq'}=\sum_{ss',kk'}\int dt''dt'''T^0(t,t'')_{pp',ss'}M(t'',t''')_{ss',kk'}T^0(t''',t')_{kk',qq'},
\label{Trans-Exp3}
\end{eqnarray}
where $T^0(t,t')_{pp',qq'}$ is given by (\ref{Trans-Unper}), and the reducible kernel $M(t,t')_{pp',qq'}$ 
takes the form
\begin{eqnarray}
M(t,t')_{pp',qq'}=
&-\sum_{n=0,m=0}&\int dt_{n-1}\cdots dt_2(\bm{S}_nG_R(t,t_{n-1})\cdots G_R(t_2,t')\bm{S}_1)_{pq}
\times
\nonumber
\\
&&\int du_1\cdots du_m(G_A(t',u_1)\bm{S}_1\cdots\bm{S}_mG_A(u_m,t))_{q'p'}
\times
\nonumber
\\
&&\text{Tr}_B[\rho_B\bm{B}_1(u_1)\cdots\bm{B}_m(u_m)\bm{B}_n(t)\cdots\bm{B}_1(t')]
\nonumber
\\
&-\sum_{n=0,m=0}&\int dt_{n}\cdots dt_1(G_R(t,t_{n})\bm{S}_n\cdots\bm{S}_1G_R(t_1,t'))_{pq}
\times
\nonumber
\\
&&\int du_2\cdots du_{m-1}(\bm{S}_1G_A(t',u_2)\cdots G_A(u_{m-1},t){\bm S_m})_{q'p'}
\times
\nonumber
\\
&&\text{Tr}_B[\rho_B{\bm B_1}(t')\cdots {\bm B_m }(t){\bm B_n}(t_n)\cdots {\bm B_1}(t_1)]
\nonumber
\\
&+\sum_{n=0,m=0}&\int dt_{n-1}\cdots dt_1({\bm S_n}G_R(t,t_{n-1})\cdots{\bm S_1} G_R(t_1,t'))_{pq}
\times
\nonumber
\\
&&\int du_2\cdots du_m({\bm S_1 }G_A(t',u_2)\cdots{\bm S_m }G_A(u_m,t))_{q'p'}
\times
\nonumber
\\
&&\text{Tr}_B[\rho_B{\bm B_1 }(t')\cdots{\bm B_m }(u_m){\bm B_n}(t)\cdots {\bm B_1}(t_1)]
\nonumber
\\
&+\sum_{n=0,m=0}&\int dt_{n}\cdots dt_2(G_R(t,t_{n}){\bm S_n}\cdots G_R(t_2,t'){\bm S_1})_{pq}
\times
\nonumber
\\
&&\int du_1\cdots du_{m-1}(G_A(t',u_1){\bm S_1 }\cdots G_A(u_{m-1},t){\bm S_m })_{q'p'}
\times
\nonumber
\\
&&\text{Tr}_B[\rho_B{\bm B_1 }(u_1)\cdots{\bm B_m}(t){\bm B_n}(t_n)\cdots {\bm B_1}(t')];
\label{M-Kern}
\end{eqnarray}
the example above applies to the third term in (\ref{M-Kern}). Inspecting Fig.~\ref{fig1} shows that 
the construction 
of the kernel $M(t,t')$ corresponds to removing the external legs $G_R(t,t_n)$, $G_R(t_1,t')$, 
$G_A(t',u_1)$, and $G_A(u_m,t)$ from the transmission matrix $T(t,t')$.

\begin{figure}
\begin{center}
\includegraphics[width=0.9\hsize]{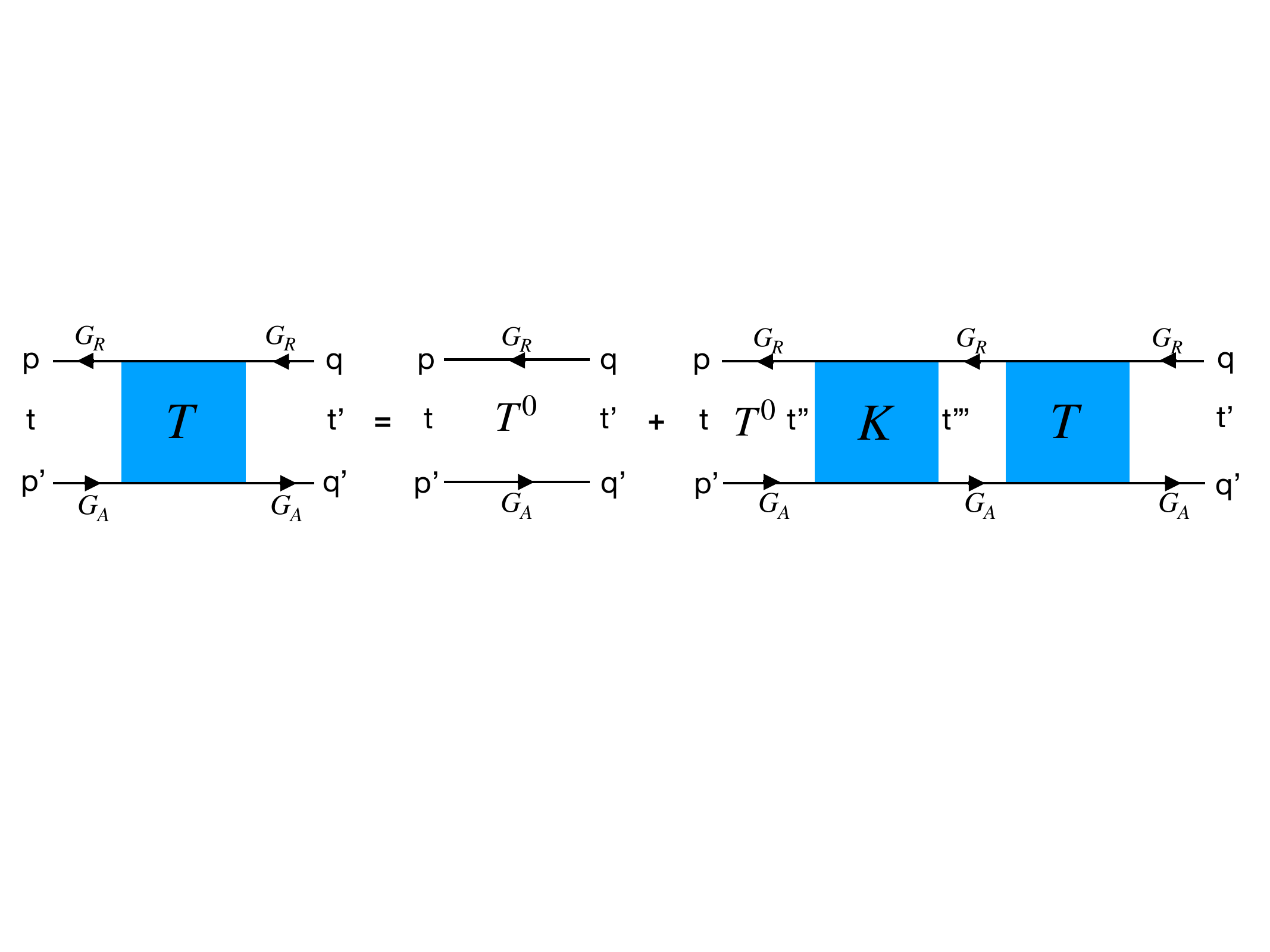}
\end{center}
\caption{Here we depict the Dyson equation for the transmission matrix. $T(t,t')$ and the irreducible kernel 
$K(t',t'')$ characterized by shaded boxes. $T^0(t,t')$ denotes the unperturbed transmission matrix, and
the retarded and advanced Green's functions $G_R(t,t')$ and $G_A(t',t)$ are denoted by directed arrows. 
}
\label{fig2}
\end{figure}

The next essential step accounting for secular effects is to identify a Dyson equation for the 
transmission matrix in (\ref{Trans-Exp}). 
According to Wick's theorem, the vertices $\bm S_p$ in the reducible kernel $M(t,t')$ in (\ref{M-Kern}) 
are connected pairwise to the bath operators $\bm B_p$ in the bath correlation function 
$D^{\alpha\beta}(t,t')$. As a result, $M(t,t')$ can 
be broken up in irreducible parts $K(t,t')$ connected by a pair of Green's functions $G_R$ and $G_A$ 
constituting the unperturbed transmission matrix $T^0(t,t')$  (\ref{Trans-Unper}). Proceeding schematically 
term by term, we have $M=I+K+KT^0K+\cdots$ and by insertion in $T=T^0MT^0$ in (\ref{Trans-Exp3})  
the expansion $T=T^0+T^0KT^0+\cdots$, yielding the Dyson equation
\begin{eqnarray}
T(t,t')_{pp',qq'}=T^0(t,t')_{pp',qq'}+\sum_{ll',ss'}\int dt''dt'''T^0(t,t'')_{pp',ll'}K(t'',t''')_{ll',ss'}T(t''',t')_{ss',qq'}.
\label{Dyson-Eq}
\end{eqnarray}
In the diagrammatic representation, the kernel $K(t,t')_{pp',qq'}$ is irreducible in the sense that is cannot 
be disconnected by the insertion of $T^0(t,t')_{pp',qq'}$. In Fig.~\ref{fig2} we have depicted the structure 
of the Dyson equation for $T(t,t')_{pp',qq'}$.
\section{\label{master} Master equation}
The Dyson equation (\ref{Dyson-Eq}) for $T(t,t')$ is a crucial results ensuring that secular effects are 
properly included. In a condensed-matter context, the irreducible kernel $K(t,t')$ serves as a 
self-energy or mass operator \cite{Abrikosov65,Mahan90,Bruus04}. Schematically (\ref{Dyson-Eq}) has the
form $T=T^0+T^0KT$. Inserted in (\ref{Trans-Mat}), $\rho_S=T\rho_S(t_i)$, we have 
$\rho_S=T^0\rho_S(t_i) +T^0KT\rho_S(t_i)$ or $\rho_S=T^0\rho_S(t_i) +T^0K\rho_S$. We thus obtain the 
following general inhomogeneous integral equation for the reduced density operator $\rho_S(t)$:
\begin{eqnarray}
\rho_S(t)_{pp'} {=} \sum_{qq'}T^0(t,t_i)_{pp',qq'}\rho_S(t_i)_{qq'}
+\sum_{ss',qq'}\int dt'dt'' T^0(t,t')_{pp',ss'}K(t',t'')_{ss',qq'}\rho_S(t'')_{qq'}.~~
\label{Integral-Equation}
\end{eqnarray}
The integral equation (\ref{Integral-Equation}) represents an integrated form of the master equation and
depends on the initial condition $\rho_S(t_i)$. 

Choosing for simplicity an energy basis, $H_S|n\rangle=E_n|n\rangle$, and using the identity 
\begin{eqnarray}
\frac{d}{dt}T^0(t,t')_{pp',qq'}=-i(E_p-E_{p'})T^0(t,t')_{pp',qq'}+\delta(t-t')\delta_{pq}\delta_{q'p'}, 
\label{Identity}
\end{eqnarray}
following from (\ref{Trans-Unper}) we infer the general master equation for $\rho_S(t)$,
\begin{eqnarray}
\frac{d}{dt}\rho_S(t)_{pp'}=-i[H_S,\rho_S(t)]_{pp'}+\sum_{qq'}\int dt'K(t,t')_{pp',qq'}\rho_S(t')_{qq'}.
\label{Master}
\end{eqnarray}
%
This is a fundamental result showing that the Dyson equation for the transmission matrix implies
a general non-Markovian master equation independent of the the initial condition.
Moreover, we note that the only assumptions underlying the structure of the master equation (\ref{Master})
are the Caldeira-Leggett heat bath in combination with Wick's theorem. In the absence of coupling
to the heat bath the kernel $K(t,t')$ vanishes and the system evolves in time according to
the von Neumann equation $d\rho_S(t)/dt=-i[H_S,\rho_S(t)]$. Since $\text{Tr}_S[\rho_S(t)]=1$,
consistency requires that the trace of $K$ vanishes, i.e., $\sum_p K(t,t')_{pp,qq'}=0$, in the ensuing
approximations.
\section{\label{born} Born approximation}
\begin{figure}
\begin{center}
\includegraphics[width=0.9\hsize]{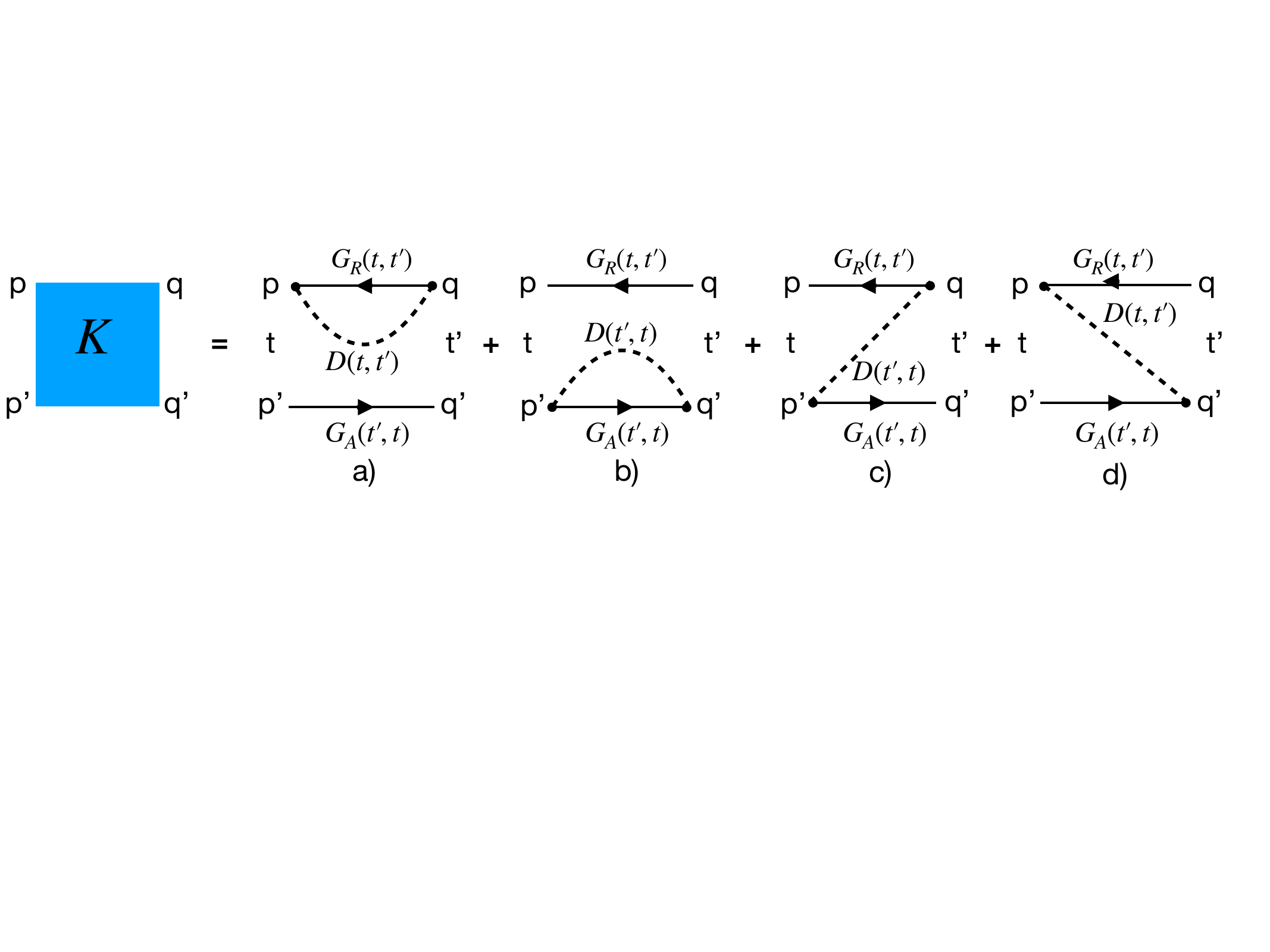}
\end{center}
\caption{Here we depict the irreducible kernel $K(t,t')$ in the Born approximation. The retarded and 
advanced Green's functions $G_R(t,t')$ and $G_A(t',t)$ are denoted by directed lines. The bath
correlation function $D(t,t')$ is denoted by a dotted line. The vertices $\bm{S}$ are denoted by dots.
Diagrams (a) and (b) correspond to populations, diagrams (c) and (d) to coherences.}
\label{fig3}
\end{figure}

The master equation in (\ref{Master}) has a general non-markovian form. In order to provide a concrete 
realization we can, in principle, expand the kernel $K(t,t')$ to any desired order in the interaction $H_{SB}$ 
by identifying the relevant diagrams and applying Wick's theorem. However, in many applications it is customary 
to assume that the coupling to the bath is weak and that it is sufficient only to consider the Born approximation, 
i.e., an expansion to second order in $H_{SB}$. In the Born approximation, only a single bath correlation 
function enters. By inspection of (\ref{M-Kern}) and inserting (\ref{Bath-Corr}) we obtain the following 
expression for the irreducible kernel $K(t,t')$:
\begin{eqnarray}
K(t,t')_{pp',qq'}=
&&-\sum_{\alpha\beta}(S^\alpha G_R(t,t')S^\beta)_{pq}G_A(t',t)_{q'p'}
D^{\alpha\beta}(t,t') 
\nonumber
\\
&&-\sum_{\alpha\beta}G_R(t,t')_{pq}(S^\alpha G_A(t',t)S^\beta)_{q'p'}
D^{\alpha\beta}(t',t)
\nonumber
\\
&&+\sum_{\alpha\beta}(S^\alpha G_R(t,t'))_{pq}(S^\beta G_A(t',t))_{q'p'}
D^{\beta\alpha}(t',t)
\nonumber
\\
&&+\sum_{\alpha\beta}(G_R(t,t')S^\alpha)_{pq}(G_A(t',t)S^{\beta})_{q'p'}
D^{\beta\alpha}(t,t').
\label{Kernel}
\end{eqnarray}
The four irreducible contributions to the kernel (\ref{Kernel}) in the Born approximation are 
depicted in Fig.~\ref{fig3}; here diagrams (a) and (b) correspond to level populations, 
whereas diagrams (c) and (d) are associated with coherences.

It is straightforward to proceed to higher order in $H_{SB}$ by identifying the corresponding
irreducible contributions to the kernel $K$. Thus to  fourth order in $H_{SB}$ one identifies
twenty individual  contributions to the kernel $K$. The issue of a stronger system-bath coupling
is  important but will not be pursued in the present context.

Inserting (\ref{Kernel}) in (\ref{Master}), we obtain the master equation 
\begin{eqnarray}
\frac{d}{dt}\rho_S(t)_{pp'}=&&-i[H_S,\rho_S(t)]_{pp'}
\nonumber
\\
&&-\int dt'\sum_{\alpha\beta, qq'}(S^\alpha G_R(t,t')S^\beta)_{pq}G_A(t',t)_{q'p'}\rho_S(t')_{qq'}
D^{\alpha\beta}(t,t') 
\nonumber
\\
&&-\int dt'\sum_{\alpha\beta,qq'}G_R(t,t')_{pq}(S^\alpha G_A(t',t)S^\beta)_{q'p'}\rho_S(t')_{qq'} 
D^{\alpha\beta}(t',t) 
\nonumber
\\
&&+\int dt'\sum_{\alpha\beta,qq'}(S^\alpha G_R(t,t'))_{pq}(S^\beta G_A(t',t))_{q'p'}\rho_S(t')_{qq'}
D^{\beta\alpha}(t',t) 
\nonumber
\\
&&+\int dt'\sum_{\alpha\beta,qq'}(G_R(t,t')S^\alpha)_{pq} (G_A(t',t)S^\beta)_{q'p'}\rho_S(t')_{qq'} 
D^{\beta\alpha}(t,t'),
\nonumber
\\
\label{Master-Born}
\end{eqnarray}
or in operator form
\begin{eqnarray}
\frac{d}{dt}\rho_S(t)=&&-i[H_S,\rho_S(t)]
\nonumber
\\
&&-\int dt'\sum_{\alpha\beta}S^\alpha G_R(t,t')S^\beta\rho_S(t')G_A(t',t)
D^{\alpha\beta}(t,t') 
\nonumber
\\
&&-\int dt'\sum_{\alpha\beta}G_R(t,t')\rho_S(t')S^\alpha G_A(t',t)S^\beta 
D^{\alpha\beta}(t',t) 
\nonumber
\\
&&+\int dt'\sum_{\alpha\beta}S^\alpha G_R(t,t')\rho_S(t')S^\beta G_A(t',t)
D^{\beta\alpha}(t',t) 
\nonumber
\\
&&+\int dt'\sum_{\alpha\beta}G_R(t,t')S^\alpha\rho_S(t')G_A(t',t)S^\beta 
D^{\beta\alpha}(t,t'). 
\label{Master-operator}
\end{eqnarray}
We already discern here the structure of the Lindblad equation in (\ref{Lindblad}). Applying the trace operation 
to (\ref{Master-operator}) and cyclically permuting the operators under the trace, we readily infer 
$d\text{Tr}_S[\rho_S(t)]/dt=0$ and, consequently, $\text{Tr}_S[\rho_S(t)]=\text{Tr}_S[\rho_S(t_i)]=1$, thus 
providing a consistency check of the Born approximation. The trace condition applied to the kernel $K$
reads $\sum_pK_{pp,qq'}=0$, which by inspection of (\ref{Kernel}) is easily verified. In terms of the diagrams
in Fig.~\ref{fig3}, the trace condition is obtained by setting $p=p'$ and noting that diagram (a) cancels with
diagram (c) and diagram (b) with diagram (d).

Introducing the Fourier transform $\tilde\rho_S(\omega)=\int dt\exp(i\omega t)\rho_s(t)$, noting that the bath 
is in a stationary state, $D^{\alpha\beta}(t,t')=D^{\alpha\beta}(t-t')$, and from 
(\ref{Retard-Green}) and (\ref{Advan-Green}) the Green's function resolvents 
$\tilde G_R(\omega)=1/(\omega-H_S+i\epsilon)$ and $\tilde G_A(\omega)=1/(\omega-H_S-i\epsilon)$,  we obtain
the master equation in Fourier space
\begin{eqnarray}
-i\omega\tilde\rho_S(\omega)_{pp'}=-iE_{pp'}\tilde\rho_S(\omega)_{pp'}+
\sum_{qq'}\tilde K(\omega)_{pp',qq'}\tilde\rho_S(\omega)_{qq'},
\label{Master-Fourier}
\end{eqnarray}
with kernel
\begin{eqnarray}
\tilde K(\omega)_{pp',qq'}=
&&-i\delta_{p'q'}\sum_{\alpha\beta,l}\int\frac{d\omega'}{2\pi}
\frac{S^\alpha_{pl}S^\beta_{lq}\delta(E_{pl}+E_{lq})
\tilde D^{\alpha\beta}(\omega')
}
{\omega-\omega'+E_{p'l}+i\epsilon}
\nonumber
\\
&&-i\delta_{pq}\sum_{\alpha\beta,l}\int\frac{d\omega'}{2\pi}
\frac{S^\alpha_{q'l}S^\beta_{lp'}\delta(E_{q'l}+E_{lp'})
\tilde D^{\alpha\beta}(-\omega')
}
{\omega-\omega'+E_{lp}+i\epsilon}
\nonumber
\\
&&+i\sum_{\alpha\beta}\int\frac{d\omega'}{2\pi}
\frac{S^\alpha_{pq}S^\beta_{q'p'}\delta(E_{pq}+E_{q'p'})
\tilde D^{\beta\alpha}(-\omega')
}
{\omega-\omega'+E_{p'q}+i\epsilon}
\nonumber
\\
&&+i\sum_{\alpha\beta}\int\frac{d\omega'}{2\pi}
\frac{S^\alpha_{pq}S^\beta_{q'p'}\delta(E_{pq}+E_{q'p'})
\tilde D^{\beta\alpha}(\omega')
}
{\omega-\omega'+E_{q'p}+i\epsilon}.
\label{Kernel-Fourier}
\end{eqnarray}
Here the matrix element of the system operator  is given by $S^\alpha_{pq}=\langle p|S^\alpha |q\rangle$ and 
the associated the energy level splitting is given by $E_{pq}=E_p-E_q$. We have, moreover, in order to implement
the rotating-wave approximation introduced in, a somewhat ad hoc manner, the delta function 
constraints
$\delta(E_{pq}+E_{q'p'})$ in order to ensure energy conservation in connection with the combined
transitions $S^\alpha_{pq}S^\beta_{q'p'}$. We note that this implementation of selection rules
does not follow automatically from the diagrammatic expansion; this is an issue that remains
to be examined. 

Applying the Plemejl formula $1/(\omega+i\epsilon)=\text{P}(1/\omega)-i\pi\delta(\omega)$ \cite{Zinn-Justin89}, 
where $\text{P}$ denotes the principal value, to the kernel (\ref{Kernel-Fourier}) we obtain, setting 
$\tilde K(\omega)_{pp',qq'}=\tilde L(\omega)_{pp',qq'}+i\tilde \Delta(\omega)_{pp',qq'}$, the  shift
\begin{eqnarray}
\tilde \Delta(\omega)_{pp',qq'}=
&&-\delta_{p'q'}\sum_{\alpha\beta,l}\text{P}\int\frac{d\omega'}{2\pi}
\frac{S^\alpha_{pl}S^\beta_{lq}\delta(E_{pl}+E_{lq})
\tilde D^{\alpha\beta}(\omega')
}
{\omega-\omega'+E_{p'l}}
\nonumber
\\
&&-\delta_{pq}\sum_{\alpha\beta,l}\text{P}\int\frac{d\omega'}{2\pi}
\frac{S^\alpha_{q'l}S^\beta_{lp'}\delta(E_{q'l}+E_{lp'})
\tilde D^{\alpha\beta}(-\omega')
}
{\omega-\omega'+E_{lp}}
\nonumber
\\
&&+\sum_{\alpha\beta}\text{P}\int\frac{d\omega'}{2\pi}
\frac{S^\alpha_{pq}S^\beta_{q'p'}\delta(E_{pq}+E_{q'p'})
\tilde D^{\beta\alpha}(-\omega')
}
{\omega-\omega'+E_{p'q}}
\nonumber
\\
&&+\sum_{\alpha\beta}\text{P}\int\frac{d\omega'}{2\pi}
\frac{S^\alpha_{pq}S^\beta_{q'p'}\delta(E_{pq}+E_{q'p'})
\tilde D^{\beta\alpha}(\omega')
}
{\omega-\omega'+E_{q'p}},
\label{Shift-Fourier}
\end{eqnarray}
and the dissipative kernel
\begin{eqnarray}
\tilde L(\omega)_{pp',qq'}=
&&-\delta_{p'q'}\frac{1}{2}\sum_{\alpha\beta,l}
S^\alpha_{pl}S^\beta_{lq}\delta(E_{pl}+E_{lq})
\tilde D^{\alpha\beta}(\omega+E_{p'l})
\nonumber
\\
&&-\delta_{pq}\frac{1}{2}\sum_{\alpha\beta,l}
S^\alpha_{q'l}S^\beta_{lp'}\delta(E_{q'l}+E_{lp'})
\tilde D^{\alpha\beta}(-\omega-E_{lp}) 
\nonumber
\\
&&+\frac{1}{2}\sum_{\alpha\beta}
S^\beta_{pq}S^\alpha_{q'p'}\delta(E_{pq}+E_{q'p'})
\tilde D^{\alpha\beta}(-\omega-E_{p'q})
\nonumber
\\
&&+\frac{1}{2}\sum_{\alpha\beta}
S^\beta_{pq}S^\alpha_{q'p'}\delta(E_{pq}+E_{q'p'})
\tilde D^{\alpha\beta}(\omega+E_{q'p}),
\label{Dissipative-Kernel-Fourier}
\end{eqnarray}
and by insertion in (\ref{Master-Fourier}) the final Fourier form of the master equation
\begin{eqnarray}
-i\omega\tilde\rho_S(\omega)_{pp'}=
-i\sum_{qq'}(E_{pp'}\delta_{pq}\delta_{p'q'}-\tilde\Delta(\omega)_{pp',qq'})\tilde \rho_S(\omega)_{qq'}
+\sum_{qq'}\tilde L(\omega)_{pp',qq'}\tilde\rho_S(\omega)_{qq'}.
\label{Final-Master-Fourier}
\end{eqnarray}
By inspection we note that $\sum_p\tilde\Delta(\omega)_{pp,qq'}=0$  and 
$\sum_p\tilde L(\omega)_{pp,qq'}=0$ yielding a vanishing trace. Moreover, we observe the general
symmetry inferred from the Dyson equation (\ref{Dyson-Eq}) together with (\ref{Trans-Unper}) or in the Born 
case from (\ref{Kernel-Fourier})
\begin{eqnarray}
&&K(t-t')_{pp',qq'}^\ast=K(t-t')_{p'p,q'q},
\\
&&\tilde K(\omega)_{pp',qq'}^\ast=\tilde K(-\omega)_{p'p,q'q}.
\label{Kernel-Symmetry}
\end{eqnarray}
The master equation (\ref{Final-Master-Fourier}) is local in Fourier space and non-markovian. The coupling of an
energy level to a continuum of states, be it the vacuum or a heat reservoir, typically gives rise to both a damping
given by $\tilde L$ and a shift given by $\tilde\Delta$. In the general case both shift and damping are 
frequency-dependent characteristic of memory effects. In the case of a Markovian behavior the shift is constant 
and appears as a Lamb shift that can be absorbed in a renormalisation of the energy levels, corresponding 
to a counter term in the system Hamiltonian $H_S$ \cite{Breuer06}
\section{\label{lindblad} Lindblad equation}
The time evolution of the density operator $\rho(t)$ for the closed system composed of the open quantum system 
under investigation and the bath is unitary and governed by the von Neumann master equation 
(\ref{vonNeumann}) with solution $\rho(t)=U(t,t_i)\rho(t_i)U(t,t_i)^\dagger$.  By construction, the density operator 
is Hermitian, positive, and has unit trace. However, due to entanglement with the bath, the reduced density 
operator $\rho_S(t)=\text{Tr}_B[\rho(t)]$ for the open quantum system does not develop in time according to a 
unitary transformation, and consequently, it does not conserve probability or entropy. On the other hand, a proper 
physical interpretation requires that $\rho_S(t)$ is Hermitian, positive, and has unit trace. These requirements 
imply that $\rho_S(t)$  conforms to the Kraus representation $\rho_S(t)=\Lambda(t,t_i)\rho_S(t_i)= 
\sum_\alpha K_\alpha(t,t_i)\rho_S(t_i)K^\dagger_\alpha(t,t_i)$, where 
$\sum_\alpha K^\dagger_\alpha K_\alpha=I$, defining  a so-called quantum channel 
\cite{Breuer06,Kraus71,Manzano20}.

In the Markov approximation, assuming a separation of the fast timescale of the bath and the slower time
scale of the open system, yielding a memory-less kernel, we have $d\rho_S(t)/dt=G\rho_S(t)$, with solution
$\rho_S(t)=\exp(G(t-t_i))\rho_S(t_i)$; here $G$ is the generator of a quantum dynamical semigroup.
The issue of the most general form of the generator $G$ has been addressed by Gorini, Kossakowski, 
Sudarshan and Lindblad (GKSL) \cite{Gorini76,Sudarshan63,Lindblad76,Manzano20}, for a review see also
\cite{Chrus17}. The GKSL or Lindblad master equation has the form given in (\ref{Lindblad}), where we note 
that the trace of the right-hand side of the equation vanishes, yielding a constant trace $\text{Tr}[\rho_S(t)]=1$.
Hermiticity, moreover, implies $\gamma_k^{\alpha\beta}=(\gamma_k^{\beta\alpha})^\ast$, i.e., 
the dissipation coefficients form a Hermitian matrix.
\section{\label{standard} Standard derivation of the Lindblad equation}
Referring for details to the standard text by Breuer and Petruccione \cite{Breuer06} (see also \cite{Manzano20}),
the customary microscopic derivation of the Lindblad equation takes as its starting point the the von Neumann 
equation in the interaction representation and its integrated form,
\begin{eqnarray}
&&\frac{d}{dt}\rho^I(t)=-i[H_{SB}(t),\rho^I(t)],
\label{L-vN}
\\
&&\rho^I(t)=\rho(0)-i\int_{0}^t dt'[H_{SB}(t'),\rho^I(t')],
\label{Int-L-vN}
\end{eqnarray}
with initial value $\rho(0)$; here $\rho^I(t)=\exp(iH_0t)\rho(t)\exp(-iH_0t)$.
Inserting (\ref{Int-L-vN}) in (\ref{L-vN}), tracing over the bath, and assuming 
$\text{Tr}_B[H_{SB}(t),\rho(t_i)]=0$, we obtain to second-order Born for
the reduced density operator $\rho^I_S(t)=\exp(iH_St)\rho_S(t)\exp(-iH_St)$
\begin{eqnarray}
\frac{d}{dt}\rho^I_S(t)=-\int_0^tdt'\text{Tr}_B[H_{SB}(t),[H_{SB}(t'),\rho^I(t')]].
\label{Int-L-vN-2}
\end{eqnarray}
Assuming a weak coupling to the reservoir and introducing the physical approximation 
\begin{eqnarray}
\rho^I(t)\approx\rho^I_S(t)\rho_B,
\label{Approx}
\end{eqnarray}
we obtain closure with respect to $\rho^I_S(t)$ yielding
\begin{eqnarray}
\frac{d}{dt}\rho^I_S(t)=-\int_0^tdt'\text{Tr}_B[H_{SB}(t),[H_{SB}(t'),\rho^I_S(t')\rho_B]].
\label{Inter}
\end{eqnarray}
We note that (\ref{Inter}) is not a proper master equation since it depends on the initial value at $t=0$;
this issue will be addressed in more detail in Sec. \ref{discussion}.

Next implementing the Markov approximation by locking $\rho^I_S(t')$ onto $\rho^I_S(t)$, 
we obtain at this stage the Redfield equation
\cite{Redfield65}
\begin{eqnarray}
\frac{d}{dt}\rho^I_S(t)=-\int_0^tdt'\text{Tr}_B[H_{SB}(t),[H_{SB}(t'),\rho^I_S(t)\rho_B]].
\label{Redfield}
\end{eqnarray}
The Redfield equation although local in time and often used in quantum optics is  
[for the same reason as (\ref{Inter})] not a full Markov equation since it depends on the initial preparation.
 However, assuming a timescale separation between
 the fast bath relaxation time $\tau_B$ and the slower system timescale $\tau_S$, i.e. $\tau_S\gg\tau_B$ 
 one obtains a Markovian master equation. In order for the resulting master equation to correspond to 
 the generator of a dynamical semi group one finally makes a further secular approximation averaging 
 over oscillating terms, the so-called rotating-wave approximation (RWA),

To implement the RWA, one projects the system operator $S^\alpha$ 
onto the energy eigenspace of the system Hamiltonian $H_S$ and defines (note that $E_{k'k}=E_{k'}-E_k$)
\begin{eqnarray}
S^\alpha(\omega)=\sum_{kk'} |k\rangle\langle k| S^\alpha |k'\rangle\langle k'| \delta(\omega-E_{k'k}).
\label{Projected-System-Operator}
\end{eqnarray}
Upon further manipulations, see \cite{Breuer06} for details, averaging over oscillating terms (the RWA)
and returning to the Schr\"{o}dinger picture, we obtain the Lindblad equation (\ref{Lindblad}) in the form
\begin{eqnarray}
&&\frac{d}{dt}\rho_S(t)=-i[H_{LS},\rho_S(t)]+ L^{\text{ST}}\rho_S(t)
\nonumber
\\
&&L^{\text{ST}}\rho_S(t)=
\sum_{\alpha\beta,\omega}\gamma_{\alpha\beta}(\omega)
\Big(S^\beta(\omega)\rho_S(t)S^{\alpha\dagger}(\omega)
-\frac{1}{2}\{S^{\alpha\dagger}(\omega)S^\beta(\omega),\rho_S(t)\}\Big).
\label{Lindblad-2}
\end{eqnarray}
Here $L^{\text{ST}}$ denotes the dissipator in the standard derivation, $H_{LS}$ is the system 
Hamiltonian including a Lamb shift, and bath correlations 
$\gamma_{\alpha\beta}(\omega)=\text{Tr}_B[\rho_BB^{\alpha\dagger} B^\beta](\omega)$.

Summarizing, the customary approach in the microscopic derivation of the Lindblad equation found
in the literature on open quantum systems
is based on a series of physical approximations: i) weak coupling to the bath, i.e. the Born approximation, 
ii) timescale separation, iii) the Markov approximation, and  iv) the RWA.
\section{\label{quasiparticle} Quasiparticle approximation}
In the master equation in the Born approximation (\ref{Master-operator}) or in the general form (\ref{Master}) the
time-dependent kernel $K(t,t')$ describes the coupling to the bath. However, the presence of memory
effects makes an analysis difficult, and it is customary to apply the Markov approximation \cite{Reichl98,Risken89}. 
This approximation is based on the assumption of a timescale separation between the fast  timescale or 
decay time of correlations in the bath and the slower timescale associated with the time evolution of the 
reduced density operator. This approach corresponds to the Langevin or equivalent Fokker-Planck scheme 
in classical statistical mechanics \cite{Risken89,Reichl98}.
\subsection{\label{qp1}Heuristic derivation of the Lindblad equation}
By inspection of (\ref{Master-operator}) we note that assuming that the bath correlations $D^{\alpha\beta}(t,t')$ 
fall off on a short timescale, $\tau_B$, compared to the timescale of the evolution of the system, $\tau_S$, 
i.e., $\tau_B\ll \tau_S$, and setting $D^{\alpha\beta}(t,t')\to \delta(t-t')D^{\alpha\beta}$ together with the 
limits $G_R(t,t)=-i(1/2)$  and $G_A(t,t)=+i(1/2)$, we recover the Lindblad equation in (\ref{Lindblad}). Note, 
however, that this heuristic argument does not provide the actual form of $D^{\alpha\beta}$ and its dependence
on the bath parameters.
\subsection{\label{qp2}Pole structure of the transmission matrix}
From (\ref{Trans-Mat}) we have in Fourier space
\begin{eqnarray}
\tilde\rho_S(\omega)_{pp'}=\sum_{qq'}\tilde T(\omega)_{pp',qq'}\exp(i\omega t_i)\rho_S(t_i)_{qq'},
\label{Trans-Mat-Fourier}
\end{eqnarray}
 and the time behavior of $\rho_S(t)_{pp'}$ is determined by the pole structure of the transmission
matrix $\tilde T(\omega)_{pp',qq'}$. From the Dyson equation (\ref{Dyson-Eq}) in Fourier space,
\begin{eqnarray}
\tilde T(\omega)_{pp',qq'}=
\tilde T^0(\omega)_{pp',qq'}+
\sum_{ss',ll'}\tilde T^0(\omega)_{pp',ss'}\tilde K(\omega)_{ss',ll'}\tilde T(\omega)_{ll',qq'},
\label{Dyson-Eq-Fourier}
\end{eqnarray}
and defining the inverse transmission matrix according to 
$\sum_{ll'}\tilde T_{pp',ll'}^{-1}\tilde T_{ll',qq'}=\delta_{pq}\delta_{p'q'}$ we infer 
\begin{eqnarray}
\tilde T(\omega)^{-1}_{pp',qq'}=\tilde T^0(\omega)^{-1}_{pp',qq'}-\tilde K(\omega)_{pp',qq'},
\label{Inverse-Dyson-Fourier}
\end{eqnarray}
yielding the resonance condition. From (\ref{Retard-Green}) and  (\ref{Advan-Green}) inserted in 
(\ref{Trans-Unper}) we have in the energy basis $\tilde T^0(\omega)_{pp',qq'}^{-1}=
-i\delta_{pq}\delta_{p'q'}(\omega-E_{pp'})$ and we obtain, splitting off the diagonal part of
$\tilde K(\omega)_{pp',qq'}$, the resonance condition given by 
\begin{eqnarray}
&&\text{det}\big[D_{pp',qq'}-\tilde K^{\text{OD}}(\omega)_{pp',qq'}\big]=0,
\\
&&D_{pp',qq'}=\big[-i(\omega-E_{pp'})-\tilde K(\omega)_{pp',pp'}\big]\delta_{pq}\delta_{p'q'},
\label{Resonance-Condition}
\end{eqnarray}
where $\tilde K^{\text{OD}}(\omega)_{pp',qq'}=(1-\delta_{pq}\delta_{p'q'})\tilde K(\omega)_{pp',qq'}$ is
the off-diagonal part (OD).
\subsection{\label{qp3}Quasiparticle approximation}
To proceed in deriving a Markov master equation, we invoke the so-called quasiparticle 
approximation employed in condensed-matter many-body theory 
\cite{Kadanoff62,Abrikosov65,Mahan90,Bruus04}. 
Here the basic building block is the single-particle Green's function $\tilde G(\omega,k)$ describing the 
propagation of a quantum particle with energy $\omega$ and momentum $k$ in a many-body environment. 
Suppressing the $k$ dependence the Green's function for a noninteracting system has the form 
$\tilde G_0(\omega)=1/(\omega-E)$, where $E$ is an energy level. In the simplest case, diagrammatic 
perturbation theory gives rise to a Dyson equation of the form
$\tilde G(\omega)=\tilde G_0(\omega)+\tilde G_0(\omega)\tilde\Sigma(\omega)\tilde G(\omega)$,
whose solution is the generic form $\tilde G(\omega)=1/[\omega-E-\tilde\Sigma(\omega)]$;
here the self-energy or mass operator $\tilde\Sigma(\omega)$ is determined perturbatively.
The time dependence of the propagation of the quasiparticle is thus given by 
the resonance condition $\tilde G(\omega)^{-1}= \omega-E-\tilde\Sigma(\omega)=0$. To leading order, 
the quasiparticle approximation corresponds to $\tilde\Sigma(\omega) \to \tilde\Sigma(E)$. Separating 
$\tilde\Sigma(E)$ in a real and imaginary part, i.e., $\tilde\Sigma(E)=\tilde\Delta+i\tilde\Gamma$, we have
$G(t)\propto e^{-i(E+\tilde\Delta)t}e^{-\tilde\Gamma t}$; the real part $\tilde\Delta$ gives rise to a 
quasiparticle energy shift, while the imaginary part $\tilde\Gamma$ yields a damping of the quasiparticle. 
Both energy shift and damping are caused by interaction with the many-body environment. For the
quasiparticle to preserve its identity, we must assume that the damping is small. It is important to note
that the Dyson equation automatically incorporates secular effects in producing an energy shift and a
damping.

Here we apply a corresponding  "quasiparticle approximation" to the transmission operator 
$\tilde T(\omega)_{pp',qq'}$ for open quantum systems in order to incorporate secular effects.
In a slightly compressed form, expressing $\text{det}[D(\omega)-\tilde K^{\text{OD}}(\omega)]$ in the form 
$\text{det}[D(\omega)]\text{det}[I-D(\omega)^{-1}\tilde K(\omega)^{\text{OD}}]$, 
using the expansion $\text{det}[I-D^{-1}\tilde K^{\text{OD}}]=1-\text{Tr}[D^{-1}\tilde K^{\text{OD}}]$, and
noting that $\text{Tr}[D^{-1}\tilde K^{\text{OD}}]=0$ (by construction), we obtain to leading order the 
resonance condition $\text{det}[D]=0$, i.e.,
\begin{eqnarray}
\text{det}\big[(-i(\omega-E_{pp'})-\tilde K(\omega)_{pp',pp'})\delta_{pq}\delta_{p'q'}\big]=0.
\label{New-Resonance-Condition}
\end{eqnarray}
For vanishing coupling for $\tilde K(\omega)_{pp',qq'}=0$ the resonance condition is given by
\begin{eqnarray}
\text{det}\big[(\omega-E_{pp'})\delta_{pq}\delta_{p'q'}\big]=0,
\label{Unper-Resonance-Condition}
\end{eqnarray}
yielding the roots $\omega=E_{pp'}$ or equivalently $\omega=E_{qq'}$ . Consequently, the 
quasiparticle approximation corresponds to replacing the frequency $\omega$ in diagonal kernel 
$\tilde K(\omega)_{pp',pp'}$ 
by the unperturbed value $\omega=E_{pp'}, E_{qq'}$. As a result, inserting in
(\ref{Shift-Fourier}) and (\ref{Dissipative-Kernel-Fourier}) we obtain for the shift and dissipative kernel in
the quasiparticle approximation the constant shift
\begin{eqnarray}
\tilde \Delta_{pp',qq'}=
&&-\delta_{p'q'}\sum_{\alpha\beta,l}\text{P}\int\frac{d\omega}{2\pi}
\frac{S^\alpha_{pl}S^\beta_{lq}\delta(E_{pl}+E_{lq})
\tilde D^{\alpha\beta}(\omega)}{E_{pl}-\omega}
\nonumber
\\
&&-\delta_{pq}\sum_{\alpha\beta,l}\text{P}\int\frac{d\omega}{2\pi}
\frac{S^\alpha_{q'l}S^\beta_{lp'}\delta(E_{q'l}+E_{lp'})
\tilde D^{\alpha\beta}(-\omega)   }{E_{lp'}-\omega}
\nonumber
\\
&&+\sum_{\alpha\beta}\text{P}\int\frac{d\omega}{2\pi}
\frac{S^\beta_{pq}S^\alpha_{q'p'}\delta(E_{pq}+E_{q'p'})
\tilde D^{\alpha\beta}(\omega)
}{E_{q'p'}-\omega}
\nonumber
\\
&&+\sum_{\alpha\beta}\text{P}\int\frac{d\omega}{2\pi}
\frac{S^\beta_{pq} S^\alpha_{q'p'}\delta(E_{pq}+E_{q'p'})
\tilde  D^{\alpha\beta}(-\omega)}{E_{pq}-\omega},
\label{Constant-Shift}
\end{eqnarray}
and the constant dissipative kernel
\begin{eqnarray}
\tilde L_{pp',qq'}=
&&-\delta_{p'q'}\frac{1}{2}\sum_{\alpha\beta,l}
S^\alpha_{pl}S^\beta_{lq}\delta(E_{pl}+E_{lq})
\tilde D^{\alpha\beta}(E_{pl})
\nonumber
\\
&&-\delta_{pq}\frac{1}{2}\sum_{\alpha\beta,l}
(S^\alpha)_{q'l}(S^\beta)_{lp'}\delta(E_{q'l}+E_{lp'})
\tilde D^{\alpha\beta}(E_{p'l})
\nonumber
\\
&&+\frac{1}{2}\sum_{\alpha\beta}
S^\beta_{pq}S^\alpha_{q'p'}\delta(E_{pq}+E_{q'p'})
\tilde D^{\alpha\beta}(E_{q'p'})
\nonumber
\\
&&+\frac{1}{2}\sum_{\alpha\beta}
S^\beta_{pq}S^\alpha_{q'p'}\delta(E_{pq}+E_{q'p'})
\tilde D^{\alpha\beta}(E_{qp}).
\label{Constant-Kernel}
\end{eqnarray}
Finally, inserting (\ref{Constant-Shift}) and (\ref{Constant-Kernel}) in (\ref{Master-Fourier}) we obtain the  
Fourier form of the master equation in the quasiparticle approximation
\begin{eqnarray}
-i\omega\tilde\rho_S(\omega)_{pp'}=
-i\sum_{qq'}(E_{pp'}\delta_{pq}\delta_{p'q'}-\tilde\Delta_{pp',qq'})\tilde \rho_S(\omega)_{qq'}
+\sum_{qq'}\tilde L_{pp',qq'}\tilde\rho_S(\omega)_{qq'}.
\label{B-QP Master-Fourier}
\end{eqnarray}
Correspondingly, the master equation takes the form
\begin{eqnarray}
\frac{d}{dt}\rho_S(t)_{pp'}=-i\sum_{qq'}(E_{pp'}\delta_{pq}\delta_{p'q'}-\tilde\Delta_{pp',qq'})\rho_S(t)_{qq'}
+\sum_{qq'}\tilde L_{pp',qq'}\rho_S(t)_{qq'}.
\label{Born-Markov-Master-Equation}
\end{eqnarray}
This is our main result, which comes from a standard field-theoretical analysis in combination with
a quasiparticle approximation and an imposed rotating-wave approximation. The master equation
is memoryless, i.e., markovian. By inspection we note that $\sum_p \tilde K_{pp,qq'}=0$, 
$\sum_p\tilde\Delta_{pp,qq'}=0$ and $\sum_p E_{pp}=0$ yielding a constant trace of $\rho_S$.

There is an important issue that we have not addressed, namely the positivity of the reduced density
matrix $\rho_S$ required from general principles, see e.g. \cite{Breuer06}. We have shown that
$\text{Tr}\rho_S=1$ both in the non-Markovian case and in the Lindblad case. The present diagrammatic
approach, however, does not ensure positivity of $\rho_S$. This issue has been discussed in 
\cite{Whitney08}.
\section{\label{discussion} Discussion}
Here we discuss the markovian Lindblad equation and non-Markovian approaches by Diosi-Feriadi and 
Nakajima-Zwanzig.
\subsection{\label{dis1}Lindblad equation}
To establish the equivalence between the present field-theoretical approach and the standard 
derivation of the Lindblad equation, we express (\ref{Lindblad-2}) in matrix form. Using
\begin{eqnarray}
&&\langle k|S^\alpha(\omega)|k'\rangle=S^\alpha_{kk'}\delta(\omega-E_{k'k}),
\\
&&\langle k|S^\alpha(\omega)^\dagger|k'\rangle=S^{\alpha\dagger}_{kk'}\delta(\omega+E_{k'k}),
\label{Matrix-Elements}
\end{eqnarray}
 summing over $\omega$, and symmetrizing the first term, we  obtain for the dissipative kernel
\begin{eqnarray}
L^{\text{ST}}_{pp',qq'}=
&&+\frac{1}{2}\sum_{\alpha\beta}
\gamma^{\alpha\beta}(E_{qp})
S^\beta_{pq}\rho_S(t)_{qq'}S^{\alpha\dagger}_{q'p'}\delta(E_{pq}+E_{q'p'})
\nonumber
\\
&&+\frac{1}{2}\sum_{\alpha\beta}
\gamma^{\alpha\beta}(E_{q'p'})
S^\beta_{pq}\rho_S(t)_{qq'}S^{\alpha\dagger}_{q'p'}\delta(E_{pq}+E_{q'p'})
\nonumber
\\
&&-\frac{1}{2}\delta_{p'q'}\sum_{\alpha\beta,l}
\gamma^{\alpha\beta}(E_{pl})
S^{\alpha\dagger}_{pl} S^\beta_{lq}\rho_S(t)_{qq'}\delta(E_{pl}+E_{lq})
\nonumber
\\
&&-\frac{1}{2}\delta_{pq}\sum_{\alpha\beta,l}
\gamma^{\alpha\beta}(E_{p'l})\rho_S(t)_{qq'}S^{\alpha\dagger}_{q'l}S^\beta_{lp'}\delta(E_{q'l}+E_{lp'}).
\label{Matrix-Standard-Lindblad}
\end{eqnarray}
Using $\sum_\alpha S^\alpha B^\alpha=\sum_\alpha S^{\alpha\dagger} B^{\alpha\dagger}$ we obtain
complete agreement with the field-theoretical expression in (\ref{Constant-Kernel}). This equivalence
demonstrates that the field-theoretical approach in combination with a RWA approximation yields the
same expression for the Lindblad equation as the standard approach. With the exception of the 
added RWA, the quasiparticle approximation replaces the physical approximation in the standard
approach.

The assumption of separation of timescales is essential in obtaining a Markov master equation and is 
used throughout in the standard derivation of the Lindblad equation. In the present field-theoretical
approach the timescale separation is implicit in the quasiparticle approximation locking the frequency
in the kernel $\tilde K(\omega)$ onto the level energy separation $\Delta E$. For this approximation
to be valid, we must assume that the kernel varies slowly over a frequency range of order 
$\tau_S\approx 1/\Delta E$. To illustrate this point, we assume that $K(t)$ due to the fast decay of the
bath correlation behaves approximately like $K(t)\approx\exp(-t/\tau_B)$, where $\tau_B$ is the 
bath correlation time. In Fourier space we then have $\tilde K(\omega)\approx 1/[\omega^2+(1/\tau_B)^2]$,
and the slow variation of $\tilde K$ implies the timescale separation $\tau_B\ll \tau_S$.
We note that a simple version of the many-body quasiparticle approximation is also encountered in the
standard Wigner-Weisskopf analysis of spontaneous emission \cite{Cohen-Tannoudji92}, where the
timescale $1/\omega_0$ associated with the level splitting $\omega_0$ is assumed to be slow compared to
the fast timescale associated with the radiation field. The resulting pole approximation corresponds to 
the Markov approximation.
\subsection{\label{dis2}Diosi - Ferialdi approach}
There is currently a strong interest in non-Markovian features of open quantum systems 
\cite{Ferialdi16,Breuer16,Breuer09,Bonifacio20,Vega17}. In this subsection we address recent 
work by Diosi and Ferialdi \cite{Diosi14,Ferialdi16}, who present an exact analytical expression for 
the transmission operator.

Choosing the initial time $t_i=0$ and inserting from Appendix \ref{app1} Eqs. (\ref{A1-16}) and (\ref{A1-17}), the
transmission operator is given by a formal expression in terms of time-ordered and anti-time-ordered
products according to
\begin{eqnarray}
T(t,0) =
&&\exp(-iH_St)\text{Tr}_B\Big(\Big[\exp\Big(-i\int_0^tdt'\bm S(t')\bm B(t')\Big)\Big]_+\rho_B\times
\nonumber
\\
&&\Big[\exp\Big(+i\int_0^tdt'\bm S(t')\bm B(t')\Big)\Big]_-\Big)\exp(+iH_St).
\nonumber
\\
\label{Exact-Transmission-1}
\end{eqnarray}
Here the time-ordered term $[\cdots]_+$ refers to the upper branch of the so-called Keldysh contour
from $t=0$ to $t$ and the anti-time-ordered term $[\cdots]_-$ to the lower branch of the Keldysh contour
from $t$ to $t=0$ \cite{Keldysh65,Schwinger61}. In the context of non equilibrium many body 
theory, the two branches are adjacent and Wick's theorem in its generator form from 
Appendix \ref{app2} (\ref{A2-7}) can be applied to the path-ordered operators along the complete 
Keldysh contour from $t=0$ to $t$ and back to $t=0$. This is the basis for diagrammatic nonequilibrium 
many body theory  \cite{Keldysh65,Schwinger61}.

In the case of an open quantum system, as exemplified in (\ref{Exact-Transmission-1}), the density operator
$\rho_B$ entering in the bath average separates the two Keldysh branches, and Wick's theorem cannot
be directly applied. In a series of intriguing papers, Diosi and Ferialdi \cite{Diosi14,Ferialdi16}, see also 
\cite{Diosi90,Diosi93},  have remedied this feature by introducing 'left' and 'right' operators according to the 
prescription $\bm S_L\bm B_L\rho_B\rho_S(0)=\bm S\bm B\rho_B\rho_S(0)$ and 
$\bm S_R\bm B_R\rho_B\rho_S(0)=\rho_B\rho_S(0)\bm S\bm B$. In this case, the equation of motion 
for the density operator in the interaction representation, $\rho^I(t)=\exp(iH_0t)\rho(t)\exp(-iH_0t)$, 
$id\rho^I(t)/dt= [\bm S(t)\bm B(t),\rho^I(t)]$, takes the form 
$id\rho^I(t)/dt= [\bm S_L(t)\bm B_L(t)-\bm S_R(t)\bm B_R(t)]\rho^I(t)$ with time-ordered solution
$\rho^I(t)=[\exp(-i\int_0^t dt'[\bm S_L(t)\bm B_L(t)-\bm S_R(t)\bm B_R(t)])]_+\rho(0)$, corresponding to
the transmission operator
\begin{eqnarray}
&&T(t,0)=
\nonumber
\\
&&\exp(-iH_0t)\text{Tr}_B\Big(\Big[\exp(-i\int_0^tdt'\Big(\bm S_L(t')\bm B_L(t')-
\bm S_R(t')\bm B_R(t')\Big)\Big]_+\rho_B\Big)\exp(+iH_0t).
\nonumber
\\
\label{Exact-Transmission-2}
\end{eqnarray}
By means of this procedure the two Keldysh branches become adjacent, and Wick's theorem can be 
applied to the complete Keldysh contour in order to explicitly average over the bath, yielding a closed 
formal expression for the transmission matrix. Referring to \cite{Diosi14,Ferialdi16} for details one arrives at
\begin{eqnarray}
&&T(t,0)=
\exp(-iH_0t)\Big[\exp\Big(\int_0^t dt'\int_0^t dt''
D^{\alpha\beta}(t',t'')Q_{RL}^{\alpha\beta}(t',t'')\Big)\Big]_+\exp(+iH_0t)
\nonumber
\\
&&Q_{RL}^{\alpha\beta}(t',t'')=S_L^\beta(t'')S_R^\alpha(t')
-\theta(t'-t'')S_L^\alpha(t')S_L^\beta(t'')-\theta(t''-t')S_R^\beta(t'')S_R^\alpha(t'),
\label{Exact-Transmission-3}
\end{eqnarray}
where the time-ordering $[\cdots]_+$ still applies to the system operators $S^{\alpha,\beta}_{R,L}(t)$ in 
$Q_{RL}^{\alpha\beta}(t',t'')$.
The result (\ref{Exact-Transmission-3}) provides an exact formal non-Markovian expression for the 
transmission operator. In further developments in \cite{Ferialdi16} Ferialdi also discusses the 
Hu-Paz-Zhang model for quantum Brownian motion \cite{Hu92}.

There is here a parallel to the Feynman path integral representation in
quantum field theory or condensed-matter many body theory  \cite{Zinn-Justin89}. In both cases, the 
closed form permits
a concise way of checking symmetries, etc. However, for practical purposes one must often resort to actual 
expansions in terms  of the interaction, typically diagrammatic expansion organised according to
appropriate Feynman rules.

In the present context of open quantum systems, the expansion of $T(t,0)$ in 
(\ref{Exact-Transmission-3}) in powers of the interaction $H_{SB}$, applying the time-order prescription 
and rearranging the system operators $S_{R,L}^{\alpha,\beta}$, should reproduce the diagrammatic expansion 
discussed here. However, we should like to emphasize that the present approach based on diagrammatic
perturbations theory allows for a derivation of the master equation, an identification of the irreducible kernel,
and diagrammatic rules for the determination of $K$ to any desired order in the interaction;
we note that the use of Wick's theorem in expanding time-ordered products is a standard tool in condensed-matter
and field theory going back to the development of quantum electrodynamics.
\subsection{\label{dis3}Nakajima - Zwanzig approach}
An approach to open quantum systems has also been formulated using the method by 
by Nakajima and Zwanzig (NZ) \cite{vanKampen92,Breuer06,Zwanzig60,Zwanzig64,Nakajima58,Butanas17,
Ivanov15,Smirne10}. This approach relies on a projection techniques yielding a formal expression for the 
master equation with a memory kernel. In the case of a memoryless kernel, the time-convolutionless 
projection operator formalism, yielding a perturbative expansion, has also been developed 
\cite{Breuer01,Chaturvedi79,Shibata77,Shibata80}. For details, we refer to \cite{Breuer06}; see also  \cite{Xu18,Venturi14,teVrugt19,teVrugt20,teVrugt21,Teretenkov19,Smirne10,Reimer19b,Nestmann21a,
Nestmann21b,Ignatyuk22}.

Below we briefly summarize the NZ approach. Referring to the exposition in \cite{Breuer06} the starting point is 
the von Neumann equation in the interaction representation in the form
\begin{eqnarray}
\frac{d\rho^I(t)}{dt}=L(t)\rho^I(t),
\label{inter-vN}
\end{eqnarray}
where $L(t)$ is the Liouville super operator acting according to $L(t)\bullet=-i[H_{SB}(t),\bullet]$;
note that $\rho^I(t)=\exp(iH_0t)\rho(t)\exp(-iH_0t)$.
The formal solution of (\ref{inter-vN}) is thus given by
\begin{eqnarray}
\rho^I(t)=\Big[\exp\Big(\int_0^tdt'L(t')\Big)\Big]_+\rho(0).
\label{NZ-sol}
\end{eqnarray}
By expanding, it is easily verified that this expression
is completely equivalent to the expression
\begin{eqnarray}
\rho^I(t)=\Big[\exp\Big(-i\int_0^tdtH_{SB}(t')\Big)\Big]_+\rho(0)\Big[\exp\Big(+i\int_0^tdt'H_{SB}(t')\Big)\Big]_-,
\label{sol2}
\end{eqnarray}
forming the basis for diagrammatic perturbation theory; here $[\cdots]_\pm$ denotes the time-ordered 
and anti-time-ordered products, respectively.

In the NZ approach, one introduces a projection operator $P$ according to the definition
\begin{eqnarray}
P\rho^I(t)=\rho^I_S(t)\rho_B,
\label{projP}
\end{eqnarray}
where $\rho^I_S(t)$ is the reduced density operator for the system, and $\rho_B$ the density operator
for the bath. Correspondingly, defining $Q=1-P$, we have the relations $P+Q=1$, $P^2=P$, $Q^2=Q$,
and $PQ=QP=0$. Applying this scheme to (\ref{inter-vN}), we obtain coupled equations of motion for 
$P\rho^I$ and $Q\rho^I$. Solving the equation for $Q\rho^I$ with initial condition $Q\rho(0)$ and inserting 
in the equation for $P\rho^I$, we obtain the Nakajima-Zwanzig equation
\begin{eqnarray}
&&\frac{dP\rho^I(t)}{dt}=PL(t)G(t,0)Q\rho(0)+PL(t)P\rho^I(t)+\int^t_0dt'K(t,t')P\rho^I(t'),
\label{NZ1}
\\
&&G(t,t')=\Big[\exp\Big(\int_{t'}^tdt'QL(t')\Big)\Big]_+,
\label{NZ-G}
\\
&&K(t,t')=PL(t)G(t,t')QL(t')P.
\label{NZ-K}
\end{eqnarray}
Assuming that odd moments of $H_{SB}$ vanish, and choosing a factorized initial condition
$\rho(0)=\rho_S(0)\rho_B$, we obtain for $P\rho(t)$
\begin{eqnarray}
&&\frac{dP\rho(t)}{dt}=\int^t_0dt'K(t,t')P\rho(t').
\label{NZ-exact}
\end{eqnarray}
To leading order in $H_{SB}$ we have $G=I$, and we obtain $K(t,t')=PL(t)QL(t')P$, yielding
the second-order NZ master equation
\begin{eqnarray}
&&\frac{dP\rho(t)}{dt}=\int^t_0dt'PL(t)L(t')P\rho(t'),
\label{NZ-exact2}
\end{eqnarray}
 which, implementing the definitions of $L$, $P$ and $Q$, agrees with the expression (\ref{Inter}) in 
Sec.  \ref{standard}.
 
 Since the projection $P\rho^I(t)=\rho^I_S(t)\rho_B$ treats $\rho_B$ as an inert background (bath) 
 and to leading order yields the expression (\ref{Inter}), it appears that  the projection basically 
 corresponds to the physical assumption $\rho^I(t)\approx\rho^I_S(t)\rho_B$ in the derivation of
 (\ref{Inter}) in Sec. \ref{standard}.
 
 Another issue regarding the NZ approach and the standard derivation in Sec. \ref{standard}
 is the dependence of the master equations (\ref{NZ-exact}), (\ref{NZ-exact2}), and (\ref{Inter})
 on the initial preparation at $t=0$, i.e., the lower integration limit. In the NZ approach, this 
 feature is associated with inserting the solution of the equation for $Q\rho^I$ with initial 
 condition $Q\rho(0)$ in the equation of motion for $P\rho^I$. Clearly, a proper evolution
 equation like the Schr\"{o}dinger equation or the von Neumann equation cannot depend
 on the initial preparation. Likewise, this must hold for a proper non-Markovian evolution 
 equation for the reduced density operator. 
 
 It seems that this dependence on the initial condition indicates that secular effects are
 not properly included in the NZ approach. In the standard derivation of the Lindblad
 equation in Sec. \ref{standard}, secular effects are included by applying the rotating-wave 
 approximation.
 
In condensed-matter many-body theory, the issue of secular effects was discussed
briefly in Sec. \ref{quasiparticle}. Secular effects are properly included by the
construction of the Dyson equation for the single-particle Green's function.
Likewise, in the present diagrammatic approach to open quantum systems, a Dyson equation
is constructed for the transmission matrix in Sec. \ref{transmission}. Schematically,
the Dyson equation for $T$ has the form $T=T^0+T^0KT$ given by (\ref{Dyson-Eq}), yielding in 
Sec. \ref{master} the general non-Markovian master equation 
$\dot\rho_S=-i[H_S,\rho_S]+K\rho_S$ in (\ref{Master}); note that if we {\em incorrectly} make a direct
expansion of the Dyson equation to leading order, i.e., $T\approx T^0+T^0KT^0$, the
definition $\rho_S=T\rho_S(0)=(T^0+T^0KT^0)\rho_S(0)$ yields a master equation
$\dot\rho_S=(\dot T^0+\dot T^0KT^0)\rho_S(0)$, depending on the initial condition at $t=0$.

The fact that the NZ approach does not include secular effects implies that we cannot compare
the NZ approach to the systematic diagrammatic method presented here. As discussed above,
even to leading order we encounter a discrepancy. Also regarding the time-convolutionless 
projection operator formalism yielding a time-local or memoryless master equation, we are 
prevented from  a direct comparison since the diagrammatic approach by construction accounts 
for memory effects.
\subsection{\label{dis4}Qubit coupled to heat bath}
To illustrate the field theoretical scheme developed in the previous sections to a particular open quantum 
system interacting with a bath, we must specify the system Hamiltonian $H_S$, the system operators 
$S^\alpha$, the bath Hamiltonian $H_B$, the bath operators $B^\alpha$, and the corresponding bath 
correlations $D^{\alpha\beta}$.  Here we consider the well-known and much studied case of a two-level 
system or qubit coupled to a single heat bath \cite{Breuer06}. The isolated qubit is characterized by the two-state
Hamiltonian 
\begin{eqnarray}
H_S=\frac{\omega_0}{2}\sigma^z,
\label{Qubit-Hamiltonian}
\end{eqnarray}
where the two energy levels are denoted  $|-\rangle$ and $|+\rangle$ with splitting $\omega_0$.
In the Pauli matrix  basis \cite{Zinn-Justin89} $\sigma^z, \sigma^\pm$ we have $\sigma^+|-\rangle=|+\rangle$, 
$\sigma^-|+\rangle=|-\rangle$, and $\sigma^z|\pm\rangle= \pm|\pm\rangle$. For the coupling to the bath we
choose $H_{SB}=\sigma^+\sum_k\lambda_k b_k+ \sigma^-\sum_k\lambda_k b_k^\dagger$. With the 
assignment $S^1=\sigma^+$, $S^2=\sigma^-$, $B^1=B$ and $B^2=B^\dagger$, the coupling is 
\begin{eqnarray}
H_{SB}=S^{1}B^{1}+S^{2}B^{2}.
\label{Qubit-Interaction-Hamiltonian}
\end{eqnarray}
By inspection of the shift (\ref{Constant-Shift}) and the kernel (\ref{Constant-Kernel}),  using 
$(S^1)_{+-}=1$, $(S^2)_{-+}=1$, and $E_{+-}=\omega_0$, and inserting $\tilde D^{12}$ 
and $\tilde D^{21}$ from (\ref{Bath1-Fourier}) and (\ref{Bath2-Fourier}), noting 
that $\tilde D^{11}=\tilde D^{22}=0$, we obtain the non vanishing shift elements
\begin{eqnarray}
&&\tilde\Delta_{+-,+-}=-\tilde\Delta_{-+,-+}=
-\text{P}\int_0\frac{d\omega}{2\pi}\frac{g(\omega)[1+2n(\omega)]}{\omega_0-\omega},
\label{Shift-Qubit}
\end{eqnarray}
and the non vanishing kernel elements
\begin{eqnarray}
&&\tilde K_{++,++}=-g(\omega_0)[1+n(\omega_0)],
\\
&&\tilde K_{--,++}=+g(\omega_0)[1+n(\omega_0)],
\\
&&\tilde K_{++,--}=+g(\omega_0)n(\omega_0),
\\
&&\tilde K_{--,--}=-g(\omega_0)n(\omega_0),
\\
&&\tilde K_{+-,+-}=-\frac{1}{2}g(\omega_0)[1+2n(\omega_0)],
\\
&&\tilde K_{-+,-+}=-\frac{1}{2}g(\omega_0)[1+2n(\omega_0)].
\label{Kernel-Qubit}
\end{eqnarray}
We note that the delta function conditions originating from the RWA are automatically satisfied
in the present case. With the notation $g(\omega_0)=g_0$, $n(\omega_0)=n_0$, and
$\tilde\Delta_{+-,+-}=\Delta$, we subsequently obtain the master equation
\begin{eqnarray}
&&\frac{d}{dt}\rho_S(t)_{++}=-g_0(1+n_0)\rho_S(t)_{++}+g_0n_0\rho_S(t)_{--},
\\
&&\frac{d}{dt}\rho_S(t)_{--}=+g_0(1+n_0)\rho_S(t)_{++}-g_0n_0\rho_S(t)_{--},
\\
&&\frac{d}{dt}\rho_S(t)_{--}=-i(\omega_0-\Delta)\rho_S(t)_{+-}-\frac{1}{2}g_0(1+2n_0)\rho_S(t)_{+-},
\\
&&\frac{d}{dt}\rho_S(t){-+}=+i(\omega_0-\Delta)\rho_S(t)_{-+}- \frac{1}{2}g_0(1+2n_0)\rho_S(t)_{-+},
\label{Master-Qubit}
\end{eqnarray}
where we note that the shift $\Delta$ can be absorbed in a renormalisation of the level shift $\omega_0$,
i.e., a Lamb shift. Finally, introducing the operators $\sigma^+$, $\sigma_-$ and $\sigma^z$, 
the corresponding Lindblad master equation has the form
\begin{eqnarray}
\frac{d}{dt}\rho_S=
-i[(\omega_0-\Delta)\sigma^z/2,\rho_S]&+&D^{-+}\bigg[\sigma^-\rho_S\sigma^+
-\frac{1}{2}\{\sigma^+\sigma^-,\rho_S\}\bigg]
\nonumber
\\
&+&D^{+-}\bigg[\sigma^+\rho_S\sigma^--\frac{1}{2}\{\sigma^-\sigma^+,\rho_S\}\bigg],
\label{Lindblad-Qubit}
\end{eqnarray}
where we have set $D^{+-}=g_0n_0$ and $D^{-+}=g_0(1+n_0)$ in compliance with (\ref{Lindblad}).

Here we discuss the quasiparticle approximation for the qubit-bath case in more detail. According to
(\ref{Inverse-Dyson-Fourier}) the inverse transmission matrix has the form
\begin{eqnarray}
&&\tilde T(\omega)^{-1}=
\nonumber
\\
&&\left(\begin{array}{cccc}
-i\omega_{++}-\tilde K_{++,++}(\omega)& -\tilde K_{++,--}(\omega)&0 &0
\\
-\tilde K_{--,++}(\omega)& -i\omega_{--}-\tilde K_{--,--}(\omega)&0 &0
\\
0&0&-i\omega_{+-}-\tilde K_{+-,+-}(\omega) &0
\\
0& 0&0 &-i\omega_{-+}-\tilde K_{-+,-+}(\omega)
\end{array}\right),
\nonumber
\\
\label{Inverse_Transmission}
\end{eqnarray}
with the notation $\omega_{pp'}=\omega-E_{pp'}$. The resonance condition is given by
$\text{det}[\tilde T(\omega)^{-1}]=0$ and we obtain 
\begin{eqnarray}
&&\omega-\omega_0+i\tilde K_{+-,+-}(\omega)=0,
\label{Det-Con1} 
\\
&&\omega+\omega_0+i\tilde K_{-+,-+}(\omega)=0,
\label{Det-Con2} 
\\
&&(\omega-i\tilde K_{++,++}(\omega))(\omega-i\tilde K_{--,--}(\omega))+
\tilde K_{++,--}(\omega)\tilde K_{--,++}(\omega)=0.
\label{Det-Con3} 
\end{eqnarray}
Applying the quasiparticle approximation we note that to leading order $\omega=\omega_0$ in (\ref{Det-Con1}),
$\omega=-\omega_0$ in (\ref{Det-Con2}), and $\omega=0$ in (\ref{Det-Con3}), yielding 
$\tilde K_{+-,+-}(\omega_0)$,
$\tilde K_{-+,-+}(-\omega_0)$, $\tilde K_{++,++}(0)$, $\tilde K_{--,--}(0)$, $\tilde K_{++,--}(0)$, 
and $\tilde K_{--,++}(0)$. Note that $K_{--,++}$ and $K_{--,--}$ also follows from the trace condition
$K_{++,++}+K_{--,++}=0$ and $K_{++,--}+K_{--,--}=0$.

\section{\label{summary} Summary}
In the present paper, we have applied condensed-matter many-body methods to
open quantum systems. We have derived a general non-Markovian master equation for the reduced density 
operator characterized by an irreducible kernel allowing for a systematic diagrammatic expansion. We have in 
particular considered the Born approximation. We have shown that the quasiparticle approximation, common in 
condensed-matter many-body theory, in the context of open quantum systems is equivalent to the standard
Markov approximation implying a separation of timescales. Implementing the rotating-wave approximation 
we have demonstrated that the Lindblad equations follows from the Markov approximation. We have,
moreover, discussed the Nakajima - Zwanzig method and its relation to the present diagrammatic
approach. As an application of the many-body approach we have discussed the coupling 
of a qubit to a thermal heat bath. Technical aspects of the 
analysis are supported by an Appendix. We believe that the present approach correctly including secular 
effects allows for a more systematic microscopic analysis of open quantum systems. 
Finally, we list here further references relevant to the present approach \cite{Reimer19,Maniscalco06,Yu06,Ignatyuk22,Schoeller09,Doll08}.
\section{\label{appendix} Appendix}
\subsection{\label{app1} Interaction Representation}
The interaction representation in the context of perturbation theory is textbook material \cite{Bruus04,Mahan90}. 
However, in order to 
render our presentation self-contained, we present the procedure below applied to open quantum systems.

The open quantum system ({\it S}) coupled to the bath ({\it B}) is described by the Hamiltonian
\begin{eqnarray}
&&H=H_S+H_B+H_{SB}, 
\label{A1-1}
\\
&&H_{SB}=\sum_\alpha S^{\alpha}B^{\alpha}=\bm{S}\cdot\bm{B},
\label{A1-2} 
\end{eqnarray}
where $H_S$ is the system Hamiltonian, $H_B$ the bath Hamiltonian, and $H_{SB}$ is the 
system-bath interaction. Here  $\bm{S}$ and $\bm{B}$ are the system and bath operators, respectively. 
The time evolution of the total system is governed by the unitary operator $U(t,t')$ satisfying 
an evolution equation, i.e.,
\begin{eqnarray}
&&U(t,t')=\exp(-iH(t-t')),
\label{A1-3}
\\
&&\frac{d}{dt}U(t,t')=-iHU(t,t'). 
\label{A1-4}
\end{eqnarray}
In order to treat the system-bath coupling $H_{SB}$ perturbatively, we apply the interaction representation. 
Introducing the Hamiltonian $H_0$ for  the uncoupled bath and system,
\begin{eqnarray}
H_0=H_S+H_B,
\label{A1-5}
\end{eqnarray}
 we have in the interaction representation
\begin{eqnarray}
&&H_{SB}(t)=\exp(iH_0t)H_{SB}\exp(-iH_0t),
\label{A1-6}
\\
&&\bm{B}(t)=\exp(iH_Bt)\bm{B}\exp(-iH_Bt),
\label{A1-7}
\\
&&\bm{S}(t)=\exp(iH_St)\bm{S}\exp(-iH_St),
\label{A1-8}
\\
&&\tilde U(t,t')=\exp(iH_0t)U(t,t')\exp(-iH_0t'),
\label{A1-9}
\end{eqnarray}
where $\tilde U(t,t')$ satisfies the evolution equation together with its integrated form,
\begin{eqnarray}
&&\frac{d}{dt}\tilde U(t,t')=-iH_{SB}(t)\tilde U(t,t'),
\label{A1-10}
\\
&&\tilde U(t,t')=I-i\int_{t'}^tdt''H_{SB}(t'')\tilde U(t'',t').
\label{A1-11}
\end{eqnarray}
Iterating (\ref{A1-11}) we obtain expansions for $\tilde U(t,t')$ and its conjugate $\tilde U(t,t')^\dagger$
according to
\begin{eqnarray}
&&\tilde U(t,t')=\sum_{n=0}(-i)^n\int_{t'}^tdt_n\int_{t'}^{t_n}dt_{n-1}\cdots\int_{t'}^{t_2}dt_1
H_{SB}(t_n)\cdots H_{SB}(t_1 ),
\label{A1-12}
\\
&&\tilde U(t,t')^\dagger=\sum_{n=0}(+i)^n\int_{t'}^tdt_n\int_{t'}^{t_n}dt_{n-1}\cdots\int_{t'}^{t_2}dt_1
H_{SB}(t_1)\cdots H_{SB}(t_n).
\label{A1-13}
\end{eqnarray}
Introducing the time-order  and anti-time-order according to the prescriptions
\begin{eqnarray}
&&(H_{SB}(t)H_{SB}(t'))_{+}=H_{SB}(t)H_{SB}(t')\eta(t-t')+H_{SB}(t')H_{SB}(t)\eta(t'-t),
\label{A1-14}
\\
&&(H_{SB}(t)H_{SB}(t'))_{-}=H_{SB}(t)H_{SB}(t')\eta(t'-t)+H_{SB}(t')H_{SB}(t)\eta(t-t'),
\label{A1-15}
\end{eqnarray}
and using (\ref{A1-9}), we have compactly for the evolution operators $U(t,t')$ and $U(t,t')^\dagger$ 
the time-ordered products \cite{Abrikosov65,Keldysh65}
\begin{eqnarray}
&&U(t,t')=\exp(-iH_0t)\Big[\exp(-i\int_{t'}^tdt''H_{SB}(t''))\Big]_+\exp(+iH_0t'),
\label{A1-16}
\\
&&U(t,t')^\dagger=\exp(-iH_0t')\Big[\exp(+i\int_{t'}^tdt''H_{SB}(t''))\Big]_-\exp(+iH_0t).
\label{A1-17}
\end{eqnarray}
Expanding (\ref{A1-14}-\ref{A1-15}) or (\ref{A1-16}-\ref{A1-17}), inserting (\ref{A1-6}-\ref{A1-8}),
and the retarded and advanced Green's functions
\begin{eqnarray}
&&G_R(t,t')=-i\eta(t-t')\exp(-iH_S(t-t')),
\label{A1-18}
\\
&&G_A(t,t')=+i\eta(t'-t)\exp(-iH_S(t-t')),
\label{A1-19}
\end{eqnarray}
we obtain the expansions
\begin{eqnarray}
&&U(t,t')=
\nonumber
\\
&&+i\sum_{n=0}\int dt_ndt_{n-1}\cdots dt_1G_R(t,t_n)
\bm{S}_{n}G_R(t_{n},t_{n-1})\bm{S}_{n-1}\cdots\bm{S}_{2}G_R(t_{2},t_{1})\bm{S}_{1}G_R(t_1,t')\times 
\nonumber
\\
&&e^{-iH_Bt}\bm{B}_{n}(t_n)\bm{B}_{n-1}(t_{n-1})\cdots\bm{B}_{2}(t_{2})\bm{B}_{1}(t_{1})e^{iH_Bt'},
\label{A1-20}
\\
&&U(t,t')^\dagger=
\nonumber
\\
&&-i\sum_{n=0}\int du_ndu_{n-1}\cdots du_1G_A(t',u_1)
\bm{S}_{1}G_A(u_{1},u_{2})\bm{S}_{2}\cdots\bm{S}_{n-1}G_A(u_{n-1},u_{n})\bm{S}_{n}G_A(u_n,t)\times 
\nonumber
\\
&&e^{-iH_Bt'}\bm{B}_{1}(u_1)\bm{B}_{2}(u_{2})\cdots\bm{B}_{n-1}(u_{n-1})\bm{B}_{n}(u_{n})e^{iH_Bt},
\label{A1-21}
\end{eqnarray}
required for the analysis in \ref{general}. 
\subsection{\label{app2} Wick's Theorem}
Within the Caldeira-Leggett prescription of the bath in terms of independent bosons (quantum oscillators) 
and assuming that the bath operators $B^\alpha(t)$ are linear combination of creation and annihilation Bose
operators, Wick's theorem implies that the average of a product of ordered bath operators 
$\text{Tr}\rho_B B_1(t_1)\cdots B_n(t_n))$ can be broken up into all possible pairings or contractions
with the time order preserved.
\subsubsection{Proof by Gaudin}
Here we summarize a proof by Gaudin \cite{Gaudin60} directly applied to a thermal
average of operator products relevant to the present analysis. Since the bath operators  $B^\alpha(t)$ are
linear combinations of the creation and annihilation operators pertaining to a specific wavenumber,
and the Hamiltonian $H_B$ is a sum of contributions from each node, it is sufficient to 
consider the thermal average of the ordered product $d_1d_2\cdots d_n$,
\begin{eqnarray}
\langle d_1d_2\cdots d_n\rangle= \text{Tr}[\rho d_1d_2\cdots d_n],
\label{A2-1}
\end{eqnarray}
with the abbreviation $d_1\equiv d_1(t_1)$. Here $d_n$ is either an annihilation operator $b_k(t)$ 
or a creation operator $b_k^\dagger(t)$ with time evolution given by (\ref{b-Evol}) and (\ref{b-dag-Evol});
from (\ref{Den-Bath}) the density operator for the k-th mode is 
$\rho=\exp(-\beta\Omega_kn_k)/\text{Tr}[\exp(-\beta\Omega_kn_k)]$. 

First cyclically moving $\rho$ to the end of the trace and subsequently permuting $d_1$ to the 
end of the operator product, we obtain,
moving the c-number commutator outside the trace, the intermediate expansion
\begin{eqnarray}
\text{Tr}[d_1d_2\cdots d_n\rho]=&&[d_1,d_2]\text{Tr}[d_3d_4\cdots d_n\rho]+\cdots
[d_1,d_n]\text{Tr}[d_2\cdots d_{n-1}\rho]
\nonumber
\\
+&&\text{Tr}[d_2\cdots d_{n}d_1\rho],
\label{A2-2}
\end{eqnarray}
Next using the identity $d_1\rho=\rho d_1z$, where $z=\exp(\beta\Omega_k)$ for 
$d_1=b_k^\dagger$ and $z=\exp(-\beta\Omega_k)$
for $d_1=b_k$, we can exchange $\rho$ and $d_1$ and by permuting the operators under the trace obtain
the expansion with the commutators $[d_1,d_p]$ replaced by $[d_1,d_p]/(1-z)$. Applying the scheme
to the case $n=2$ we infer
\begin{eqnarray}
[d_1,d_2]/(1-z)=\text{Tr}[\rho d_1d_2]=\langle d_1d_2 \rangle.
\label{A2-3}
\end{eqnarray}
Finally, we have
\begin{eqnarray}
\langle d_1d_2\cdots d_n\rangle=\langle d_1d_2 \rangle\langle d_3d_4\cdots d_n \rangle+\cdots
\langle d_1d_n \rangle\langle d_2\cdots d_{n-1}\rangle,
\label{A2-4}
\end{eqnarray}
and by induction Wick's theorem, i.e., the ordered average is reduced to all possible pairings of 
two operators, where we note that the order is preserved as shown in (\ref{Wick-Four}).
\subsubsection{Wick's Theorem in generator form}
In the expression (\ref{A2-1}) we have assumed a specific order of the operators $\{d_n(t)\}$; however, introducing
the time order and anti-time order prescriptions
\begin{eqnarray}
&&[d(t_n)d(t_m)]_+=d(t_n)d(t_m)\theta(t_n-t_m)+d_md_n\theta(t_m-t_n),
\label{A2-5}
\\
&&[d(t_n)d(t_m)]_-=d(t_n)d(t_m)\theta(t_m-t_n)+d(t_m)d(t_n)\theta(t_n-t_m),
\label{A2-6}
\end{eqnarray}
and noting that the operators commute under time ordering or anti-time ordering, we infer the functional
Wick theorem \cite{Chou85,Zinn-Justin89}
\begin{eqnarray}
\Big\langle\Big[\exp\Big(-i\int dt\Omega(t)d(t)\Big)\Big]_\pm\Big\rangle=
\exp\Big(-\frac{1}{2}\int dtdt'\Omega(t)\Big\langle\Big[d(t)d(t')\Big]_\pm\Big\rangle\Omega(t')\Big),
\label{A2-7}
\end{eqnarray}
where $\Omega(t)$ is a generator field. 
%
%


\clearpage
\newpage
\begin{thebibliography}{94}
\expandafter\ifx\csname natexlab\endcsname\relax\def\natexlab#1{#1}\fi
\expandafter\ifx\csname bibnamefont\endcsname\relax
  \def\bibnamefont#1{#1}\fi
\expandafter\ifx\csname bibfnamefont\endcsname\relax
  \def\bibfnamefont#1{#1}\fi
\expandafter\ifx\csname citenamefont\endcsname\relax
  \def\citenamefont#1{#1}\fi
\expandafter\ifx\csname url\endcsname\relax
  \def\url#1{\texttt{#1}}\fi
\expandafter\ifx\csname urlprefix\endcsname\relax\def\urlprefix{URL }\fi
\providecommand{\bibinfo}[2]{#2}
\providecommand{\eprint}[2][]{\url{#2}}

\bibitem[{\citenamefont{Scully and Zubairy}(1996)}]{Scully96}
\bibinfo{author}{\bibfnamefont{M.}~\bibnamefont{Scully}} \bibnamefont{and}
  \bibinfo{author}{\bibfnamefont{M.~S.} \bibnamefont{Zubairy}},
  \emph{\bibinfo{title}{Quantum Optics}} (\bibinfo{publisher}{Akademic},
  \bibinfo{address}{Berlin}, \bibinfo{year}{1996}).

\bibitem[{\citenamefont{Kampen}(1992)}]{vanKampen92}
\bibinfo{author}{\bibfnamefont{N.~G.~V.} \bibnamefont{Kampen}},
  \emph{\bibinfo{title}{Stochastic Processes in Physics and Chemistry}}
  (\bibinfo{publisher}{North-Holland}, \bibinfo{address}{Amsterdam},
  \bibinfo{year}{1992}).

\bibitem[{\citenamefont{Nielsen and Chuang}(2010)}]{Nielsen10}
\bibinfo{author}{\bibfnamefont{M.~A.} \bibnamefont{Nielsen}} \bibnamefont{and}
  \bibinfo{author}{\bibfnamefont{I.~L.} \bibnamefont{Chuang}},
  \emph{\bibinfo{title}{Quantum Computation and Quantum Information: 10th
  Anniversary Edition}} (\bibinfo{publisher}{Cambridge University Press},
  \bibinfo{address}{Cambridge}, \bibinfo{year}{2010}).

\bibitem[{\citenamefont{Feshbach}(1958)}]{Feshbach58}
\bibinfo{author}{\bibfnamefont{H.}~\bibnamefont{Feshbach}},
  \bibinfo{journal}{Ann. Phys.} \textbf{\bibinfo{volume}{5}},
  \bibinfo{pages}{357} (\bibinfo{year}{1958}).

\bibitem[{\citenamefont{Zanardi and Rasetti}(1997)}]{Zanardi97}
\bibinfo{author}{\bibfnamefont{P.}~\bibnamefont{Zanardi}} \bibnamefont{and}
  \bibinfo{author}{\bibfnamefont{M.}~\bibnamefont{Rasetti}},
  \bibinfo{journal}{Phys. Rev. Lett.} \textbf{\bibinfo{volume}{79}},
  \bibinfo{pages}{3306} (\bibinfo{year}{1997}).

\bibitem[{\citenamefont{Bourennane et~al.}(2004)\citenamefont{Bourennane, Eibl,
  Gaertner, Kurtsiefer, A.Cabello, and Weinfurter}}]{Bourennane04}
\bibinfo{author}{\bibfnamefont{M.}~\bibnamefont{Bourennane}},
  \bibinfo{author}{\bibfnamefont{M.}~\bibnamefont{Eibl}},
  \bibinfo{author}{\bibfnamefont{S.}~\bibnamefont{Gaertner}},
  \bibinfo{author}{\bibfnamefont{C.}~\bibnamefont{Kurtsiefer}},
  \bibinfo{author}{\bibnamefont{A.Cabello}}, \bibnamefont{and}
  \bibinfo{author}{\bibfnamefont{H.}~\bibnamefont{Weinfurter}},
  \bibinfo{journal}{Phys. Rev. Lett.} \textbf{\bibinfo{volume}{92}},
  \bibinfo{pages}{107901} (\bibinfo{year}{2004}).

\bibitem[{\citenamefont{Verstraete et~al.}(2009)\citenamefont{Verstraete, Wolf,
  and Cirac}}]{Verstraete09}
\bibinfo{author}{\bibfnamefont{F.}~\bibnamefont{Verstraete}},
  \bibinfo{author}{\bibfnamefont{M.~M.} \bibnamefont{Wolf}}, \bibnamefont{and}
  \bibinfo{author}{\bibfnamefont{J.~I.} \bibnamefont{Cirac}},
  \bibinfo{journal}{Nature Physics} \textbf{\bibinfo{volume}{5}},
  \bibinfo{pages}{633} (\bibinfo{year}{2009}).

\bibitem[{\citenamefont{Diehl et~al.}(2011)\citenamefont{Diehl, Rico, Baranov,
  and Zoller}}]{Diehl11}
\bibinfo{author}{\bibfnamefont{S.}~\bibnamefont{Diehl}},
  \bibinfo{author}{\bibfnamefont{E.}~\bibnamefont{Rico}},
  \bibinfo{author}{\bibfnamefont{M.~A.} \bibnamefont{Baranov}},
  \bibnamefont{and} \bibinfo{author}{\bibfnamefont{P.}~\bibnamefont{Zoller}},
  \bibinfo{journal}{Nature Physics} \textbf{\bibinfo{volume}{7}},
  \bibinfo{pages}{971} (\bibinfo{year}{2011}).

\bibitem[{\citenamefont{Breuer and Petruccione}(2006)}]{Breuer06}
\bibinfo{author}{\bibfnamefont{H.~P.} \bibnamefont{Breuer}} \bibnamefont{and}
  \bibinfo{author}{\bibfnamefont{F.}~\bibnamefont{Petruccione}},
  \emph{\bibinfo{title}{The Theory of Open Quantum Systems}}
  (\bibinfo{publisher}{Oxford University Press}, \bibinfo{address}{Oxford},
  \bibinfo{year}{2006}).

\bibitem[{\citenamefont{Walls and Milburn}(1994)}]{Walls94}
\bibinfo{author}{\bibfnamefont{D.~F.} \bibnamefont{Walls}} \bibnamefont{and}
  \bibinfo{author}{\bibfnamefont{G.~J.} \bibnamefont{Milburn}},
  \emph{\bibinfo{title}{Quantum Optics}} (\bibinfo{publisher}{Springer-Verlag},
  \bibinfo{address}{New York}, \bibinfo{year}{1994}).

\bibitem[{\citenamefont{Gardiner and Zoller}(2005)}]{Gardiner05}
\bibinfo{author}{\bibfnamefont{C.~W.} \bibnamefont{Gardiner}} \bibnamefont{and}
  \bibinfo{author}{\bibfnamefont{P.}~\bibnamefont{Zoller}},
  \emph{\bibinfo{title}{Quantum Noise}} (\bibinfo{publisher}{Springer-Verlag},
  \bibinfo{address}{Berlin}, \bibinfo{year}{2005}).

\bibitem[{\citenamefont{Aolita et~al.}(2015)\citenamefont{Aolita, de~Melo, and
  Davidovich}}]{Aolita15}
\bibinfo{author}{\bibfnamefont{L.}~\bibnamefont{Aolita}},
  \bibinfo{author}{\bibfnamefont{F.}~\bibnamefont{de~Melo}}, \bibnamefont{and}
  \bibinfo{author}{\bibfnamefont{L.}~\bibnamefont{Davidovich}},
  \bibinfo{journal}{Rep. Prog. Phys.} \textbf{\bibinfo{volume}{78}},
  \bibinfo{pages}{042001 (79pp)} (\bibinfo{year}{2015}).

\bibitem[{\citenamefont{Bellomo et~al.}(2007)\citenamefont{Bellomo, LoFranco,
  and Compagno}}]{Bellomo07}
\bibinfo{author}{\bibfnamefont{B.}~\bibnamefont{Bellomo}},
  \bibinfo{author}{\bibfnamefont{R.}~\bibnamefont{LoFranco}}, \bibnamefont{and}
  \bibinfo{author}{\bibfnamefont{G.}~\bibnamefont{Compagno}},
  \bibinfo{journal}{Phys. Rev. Lett.} \textbf{\bibinfo{volume}{99}},
  \bibinfo{pages}{160502} (\bibinfo{year}{2007}).

\bibitem[{\citenamefont{Schlosshauer}(2007)}]{Schlosshauer07}
\bibinfo{author}{\bibfnamefont{M.~A.} \bibnamefont{Schlosshauer}},
  \emph{\bibinfo{title}{Decoherence and the Quantum-To-Classical Transition}}
  (\bibinfo{publisher}{Springer-Verlag}, \bibinfo{address}{Berlin},
  \bibinfo{year}{2007}).

\bibitem[{\citenamefont{Schlosshauer}(2019)}]{Schlosshauer19}
\bibinfo{author}{\bibfnamefont{M.~A.} \bibnamefont{Schlosshauer}},
  \bibinfo{journal}{Physics Reports} \textbf{\bibinfo{volume}{831}},
  \bibinfo{pages}{1} (\bibinfo{year}{2019}).

\bibitem[{\citenamefont{Paneru et~al.}(2020)\citenamefont{Paneru, Cohen,
  Fickler, Boyd, and Karimi}}]{Paneru20}
\bibinfo{author}{\bibfnamefont{D.}~\bibnamefont{Paneru}},
  \bibinfo{author}{\bibfnamefont{E.}~\bibnamefont{Cohen}},
  \bibinfo{author}{\bibfnamefont{R.}~\bibnamefont{Fickler}},
  \bibinfo{author}{\bibfnamefont{R.~W.} \bibnamefont{Boyd}}, \bibnamefont{and}
  \bibinfo{author}{\bibfnamefont{E.}~\bibnamefont{Karimi}},
  \bibinfo{journal}{Rep. Prog. Phys.} \textbf{\bibinfo{volume}{83}},
  \bibinfo{pages}{064001 (19pp)} (\bibinfo{year}{2020}).

\bibitem[{\citenamefont{Goold et~al.}(2016)\citenamefont{Goold, Huber, Riera,
  del Rio, and Skrzypczyk}}]{Goold16}
\bibinfo{author}{\bibfnamefont{J.}~\bibnamefont{Goold}},
  \bibinfo{author}{\bibfnamefont{M.}~\bibnamefont{Huber}},
  \bibinfo{author}{\bibfnamefont{A.}~\bibnamefont{Riera}},
  \bibinfo{author}{\bibfnamefont{L.}~\bibnamefont{del Rio}}, \bibnamefont{and}
  \bibinfo{author}{\bibfnamefont{P.}~\bibnamefont{Skrzypczyk}},
  \bibinfo{journal}{J. Phys. A: Math. Theor.} \textbf{\bibinfo{volume}{49}},
  \bibinfo{pages}{143001 (50pp)} (\bibinfo{year}{2016}).

\bibitem[{\citenamefont{Davidovich}(2016)}]{Davidovich16}
\bibinfo{author}{\bibfnamefont{L.}~\bibnamefont{Davidovich}},
  \bibinfo{journal}{Phys. Scr.} \textbf{\bibinfo{volume}{91}},
  \bibinfo{pages}{063013 (9pp)} (\bibinfo{year}{2016}).

\bibitem[{\citenamefont{Nakajima}(1958)}]{Nakajima58}
\bibinfo{author}{\bibfnamefont{S.}~\bibnamefont{Nakajima}},
  \bibinfo{journal}{Progress of Theoretical Physics}
  \textbf{\bibinfo{volume}{20}}, \bibinfo{pages}{948} (\bibinfo{year}{1958}).

\bibitem[{\citenamefont{Wangsness and Bloch}(1953)}]{Wangsness53}
\bibinfo{author}{\bibfnamefont{R.~K.} \bibnamefont{Wangsness}}
  \bibnamefont{and} \bibinfo{author}{\bibfnamefont{F.}~\bibnamefont{Bloch}},
  \bibinfo{journal}{Phys. Rev. 89} \textbf{\bibinfo{volume}{89}},
  \bibinfo{pages}{728} (\bibinfo{year}{1953}).

\bibitem[{\citenamefont{Redfield}(1965)}]{Redfield65}
\bibinfo{author}{\bibfnamefont{A.~G.} \bibnamefont{Redfield}},
  \bibinfo{journal}{Advances in Magnetic and Optical Resonance}
  \textbf{\bibinfo{volume}{1}}, \bibinfo{pages}{1} (\bibinfo{year}{1965}).

\bibitem[{\citenamefont{Davies}(1974)}]{Davies74}
\bibinfo{author}{\bibfnamefont{E.~B.} \bibnamefont{Davies}},
  \bibinfo{journal}{Commun. Math. Phys.} \textbf{\bibinfo{volume}{39}},
  \bibinfo{pages}{91} (\bibinfo{year}{1974}).

\bibitem[{\citenamefont{Majenz et~al.}(2013)\citenamefont{Majenz, Albash,
  Breuer, and Lidar}}]{Majenz13}
\bibinfo{author}{\bibfnamefont{C.}~\bibnamefont{Majenz}},
  \bibinfo{author}{\bibfnamefont{T.}~\bibnamefont{Albash}},
  \bibinfo{author}{\bibfnamefont{H.-P.} \bibnamefont{Breuer}},
  \bibnamefont{and} \bibinfo{author}{\bibfnamefont{D.~A.} \bibnamefont{Lidar}},
  \bibinfo{journal}{Phys. Rev. A} \textbf{\bibinfo{volume}{88}},
  \bibinfo{pages}{012103} (\bibinfo{year}{2013}).

\bibitem[{\citenamefont{Mozgunov and Lidar}(2020)}]{Mozgunov20}
\bibinfo{author}{\bibfnamefont{E.}~\bibnamefont{Mozgunov}} \bibnamefont{and}
  \bibinfo{author}{\bibfnamefont{D.}~\bibnamefont{Lidar}},
  \bibinfo{journal}{Quantum} \textbf{\bibinfo{volume}{4}}, \bibinfo{pages}{227}
  (\bibinfo{year}{2020}).

\bibitem[{\citenamefont{Rivas and Huelga}(2012)}]{Rivas12}
\bibinfo{author}{\bibfnamefont{A.}~\bibnamefont{Rivas}} \bibnamefont{and}
  \bibinfo{author}{\bibfnamefont{S.~F.} \bibnamefont{Huelga}},
  \emph{\bibinfo{title}{Open Quantum Systems}}
  (\bibinfo{publisher}{Springer-Verlag}, \bibinfo{address}{Berlin},
  \bibinfo{year}{2012}).

\bibitem[{\citenamefont{Plenio and Knight}(1998)}]{Plenio98}
\bibinfo{author}{\bibfnamefont{M.~B.} \bibnamefont{Plenio}} \bibnamefont{and}
  \bibinfo{author}{\bibfnamefont{P.~L.} \bibnamefont{Knight}},
  \bibinfo{journal}{Rev. Mod. Phys.} \textbf{\bibinfo{volume}{70}},
  \bibinfo{pages}{101} (\bibinfo{year}{1998}).

\bibitem[{\citenamefont{Carmichael}(1991)}]{Carmichael91}
\bibinfo{author}{\bibfnamefont{H.~J.} \bibnamefont{Carmichael}},
  \emph{\bibinfo{title}{An Open Systems Approach to Quantum Optics}}
  (\bibinfo{publisher}{Springer-Verlag}, \bibinfo{address}{Berlin},
  \bibinfo{year}{1991}).

\bibitem[{\citenamefont{Dalibard et~al.}(1992)\citenamefont{Dalibard, Castin,
  and M{\o}lmer}}]{Dalibard92}
\bibinfo{author}{\bibfnamefont{J.}~\bibnamefont{Dalibard}},
  \bibinfo{author}{\bibfnamefont{Y.}~\bibnamefont{Castin}}, \bibnamefont{and}
  \bibinfo{author}{\bibfnamefont{K.}~\bibnamefont{M{\o}lmer}},
  \bibinfo{journal}{Phys. Rev. Lett.} \textbf{\bibinfo{volume}{68}},
  \bibinfo{pages}{580} (\bibinfo{year}{1992}).

\bibitem[{\citenamefont{M{\o}lmer et~al.}(1993)\citenamefont{M{\o}lmer, Castin,
  and Dalibard}}]{Moelmer93}
\bibinfo{author}{\bibfnamefont{K.}~\bibnamefont{M{\o}lmer}},
  \bibinfo{author}{\bibfnamefont{Y.}~\bibnamefont{Castin}}, \bibnamefont{and}
  \bibinfo{author}{\bibfnamefont{J.}~\bibnamefont{Dalibard}},
  \bibinfo{journal}{JOSA B} \textbf{\bibinfo{volume}{10}}, \bibinfo{pages}{524}
  (\bibinfo{year}{1993}).

\bibitem[{\citenamefont{Rammer}(1991)}]{Rammer91}
\bibinfo{author}{\bibfnamefont{J.}~\bibnamefont{Rammer}},
  \bibinfo{journal}{Rev. Mod. Phys.} \textbf{\bibinfo{volume}{63}},
  \bibinfo{pages}{781} (\bibinfo{year}{1991}).

\bibitem[{\citenamefont{Fogedby}(1993)}]{Fogedby93}
\bibinfo{author}{\bibfnamefont{H.~C.} \bibnamefont{Fogedby}},
  \bibinfo{journal}{Phys. Rev. A} \textbf{\bibinfo{volume}{47}},
  \bibinfo{pages}{4364} (\bibinfo{year}{1993}).

\bibitem[{\citenamefont{Mahan}(1990)}]{Mahan90}
\bibinfo{author}{\bibfnamefont{G.~D.} \bibnamefont{Mahan}},
  \emph{\bibinfo{title}{Many Particle Physics}} (\bibinfo{publisher}{Plenum
  Press}, \bibinfo{address}{New York}, \bibinfo{year}{1990}).

\bibitem[{\citenamefont{Bruus and Flensberg}(2004)}]{Bruus04}
\bibinfo{author}{\bibfnamefont{H.}~\bibnamefont{Bruus}} \bibnamefont{and}
  \bibinfo{author}{\bibfnamefont{K.}~\bibnamefont{Flensberg}},
  \emph{\bibinfo{title}{Many-Body Quantum Theory in Condensed Matter Physics:
  An Introduction}} (\bibinfo{publisher}{Oxford University Press},
  \bibinfo{address}{Oxford}, \bibinfo{year}{2004}).

\bibitem[{\citenamefont{Landau and Lifshitz}(1980)}]{Landau80}
\bibinfo{author}{\bibfnamefont{L.}~\bibnamefont{Landau}} \bibnamefont{and}
  \bibinfo{author}{\bibfnamefont{E.}~\bibnamefont{Lifshitz}},
  \emph{\bibinfo{title}{Statistical Physics}} (\bibinfo{publisher}{Pergamon
  Press}, \bibinfo{address}{Oxford}, \bibinfo{year}{1980}).

\bibitem[{\citenamefont{Reichl}(1998)}]{Reichl98}
\bibinfo{author}{\bibfnamefont{L.~E.} \bibnamefont{Reichl}},
  \emph{\bibinfo{title}{A Modern Course in Statistical Physics}}
  (\bibinfo{publisher}{Wiley}, \bibinfo{address}{New York},
  \bibinfo{year}{1998}).

\bibitem[{\citenamefont{Alipour et~al.}(2020)\citenamefont{Alipour, Rezakhani,
  Babu, M{\o}lmer, M{\"o}tt{\"o}nen, and Ala-Nissila}}]{Alipour20}
\bibinfo{author}{\bibfnamefont{S.}~\bibnamefont{Alipour}},
  \bibinfo{author}{\bibfnamefont{A.~T.} \bibnamefont{Rezakhani}},
  \bibinfo{author}{\bibfnamefont{A.~P.} \bibnamefont{Babu}},
  \bibinfo{author}{\bibfnamefont{K.}~\bibnamefont{M{\o}lmer}},
  \bibinfo{author}{\bibfnamefont{M.}~\bibnamefont{M{\"o}tt{\"o}nen}},
  \bibnamefont{and}
  \bibinfo{author}{\bibfnamefont{T.}~\bibnamefont{Ala-Nissila}},
  \bibinfo{journal}{Phys. Rev. X} \textbf{\bibinfo{volume}{10}},
  \bibinfo{pages}{041024} (\bibinfo{year}{2020}).

\bibitem[{\citenamefont{Caldeira and Leggett}(1983)}]{Caldeira83a}
\bibinfo{author}{\bibfnamefont{A.~O.} \bibnamefont{Caldeira}} \bibnamefont{and}
  \bibinfo{author}{\bibfnamefont{A.~J.} \bibnamefont{Leggett}},
  \bibinfo{journal}{Ann. Phys.} \textbf{\bibinfo{volume}{149}},
  \bibinfo{pages}{374} (\bibinfo{year}{1983}).

\bibitem[{\citenamefont{Caldeira}(1983)}]{Caldeira83b}
\bibinfo{author}{\bibfnamefont{A.~O.} \bibnamefont{Caldeira}},
  \bibinfo{journal}{Physica} \textbf{\bibinfo{volume}{121A}},
  \bibinfo{pages}{587} (\bibinfo{year}{1983}).

\bibitem[{\citenamefont{Landau and Lifshitz}(1959)}]{Landau59}
\bibinfo{author}{\bibfnamefont{L.}~\bibnamefont{Landau}} \bibnamefont{and}
  \bibinfo{author}{\bibfnamefont{E.}~\bibnamefont{Lifshitz}},
  \emph{\bibinfo{title}{Quantum Mechanics}} (\bibinfo{publisher}{Pergamon
  Press}, \bibinfo{address}{Oxford}, \bibinfo{year}{1959}).

\bibitem[{\citenamefont{Risken}(1989)}]{Risken89}
\bibinfo{author}{\bibfnamefont{H.}~\bibnamefont{Risken}},
  \emph{\bibinfo{title}{The Fokker-Planck Equation}}
  (\bibinfo{publisher}{Springer-Verlag}, \bibinfo{address}{Berlin},
  \bibinfo{year}{1989}).

\bibitem[{\citenamefont{Gorini et~al.}(1978)\citenamefont{Gorini, Fricero,
  Verri, Kossakowski, and Sudarshan}}]{Gorini78}
\bibinfo{author}{\bibfnamefont{V.}~\bibnamefont{Gorini}},
  \bibinfo{author}{\bibfnamefont{A.}~\bibnamefont{Fricero}},
  \bibinfo{author}{\bibfnamefont{M.}~\bibnamefont{Verri}},
  \bibinfo{author}{\bibfnamefont{A.}~\bibnamefont{Kossakowski}},
  \bibnamefont{and} \bibinfo{author}{\bibfnamefont{E.~C.~G.}
  \bibnamefont{Sudarshan}}, \bibinfo{journal}{Reports on Mathematical Physics}
  \textbf{\bibinfo{volume}{13}}, \bibinfo{pages}{149} (\bibinfo{year}{1978}).

\bibitem[{\citenamefont{Diosi and Ferialdi}(2014)}]{Diosi14}
\bibinfo{author}{\bibfnamefont{L.}~\bibnamefont{Diosi}} \bibnamefont{and}
  \bibinfo{author}{\bibfnamefont{L.}~\bibnamefont{Ferialdi}},
  \bibinfo{journal}{Phys. Rev. Lett.} \textbf{\bibinfo{volume}{113}},
  \bibinfo{pages}{200403} (\bibinfo{year}{2014}).

\bibitem[{\citenamefont{Ferialdi}(2016)}]{Ferialdi16}
\bibinfo{author}{\bibfnamefont{L.}~\bibnamefont{Ferialdi}},
  \bibinfo{journal}{Phys. Rev. Lett} \textbf{\bibinfo{volume}{116}},
  \bibinfo{pages}{120402} (\bibinfo{year}{2016}).

\bibitem[{\citenamefont{Breuer et~al.}(2016)\citenamefont{Breuer, Laine, Piilo,
  and Vacchini}}]{Breuer16}
\bibinfo{author}{\bibfnamefont{H.~P.} \bibnamefont{Breuer}},
  \bibinfo{author}{\bibfnamefont{E.-M.} \bibnamefont{Laine}},
  \bibinfo{author}{\bibfnamefont{J.}~\bibnamefont{Piilo}}, \bibnamefont{and}
  \bibinfo{author}{\bibfnamefont{B.}~\bibnamefont{Vacchini}},
  \bibinfo{journal}{Rev. Mod. Phys.} \textbf{\bibinfo{volume}{88}},
  \bibinfo{pages}{021002} (\bibinfo{year}{2016}).

\bibitem[{\citenamefont{Breuer et~al.}(2009)\citenamefont{Breuer, Laine, and
  Piilo}}]{Breuer09}
\bibinfo{author}{\bibfnamefont{H.~P.} \bibnamefont{Breuer}},
  \bibinfo{author}{\bibfnamefont{E.-M.} \bibnamefont{Laine}}, \bibnamefont{and}
  \bibinfo{author}{\bibfnamefont{J.}~\bibnamefont{Piilo}},
  \bibinfo{journal}{Phys. Rev. Lett.} \textbf{\bibinfo{volume}{103}},
  \bibinfo{pages}{210401} (\bibinfo{year}{2009}).

\bibitem[{\citenamefont{Bonifacio and Budini}(2020)}]{Bonifacio20}
\bibinfo{author}{\bibfnamefont{M.}~\bibnamefont{Bonifacio}} \bibnamefont{and}
  \bibinfo{author}{\bibfnamefont{A.~A.} \bibnamefont{Budini}},
  \bibinfo{journal}{Phys. Rev. A} \textbf{\bibinfo{volume}{102}},
  \bibinfo{pages}{022216} (\bibinfo{year}{2020}).

\bibitem[{\citenamefont{de~Vega and Alonso}(2017)}]{Vega17}
\bibinfo{author}{\bibfnamefont{I.}~\bibnamefont{de~Vega}} \bibnamefont{and}
  \bibinfo{author}{\bibfnamefont{D.}~\bibnamefont{Alonso}},
  \bibinfo{journal}{Rev. Mod. Phys.} \textbf{\bibinfo{volume}{89}},
  \bibinfo{pages}{015001} (\bibinfo{year}{2017}).

\bibitem[{\citenamefont{Lindblad}(1976)}]{Lindblad76}
\bibinfo{author}{\bibfnamefont{G.}~\bibnamefont{Lindblad}},
  \bibinfo{journal}{Commun. Math. Phys.} \textbf{\bibinfo{volume}{48}},
  \bibinfo{pages}{119} (\bibinfo{year}{1976}).

\bibitem[{\citenamefont{Manzano}(2020)}]{Manzano20}
\bibinfo{author}{\bibfnamefont{D.}~\bibnamefont{Manzano}},
  \bibinfo{journal}{AIP Advances 10} p. \bibinfo{pages}{025106}
  (\bibinfo{year}{2020}).

\bibitem[{\citenamefont{Abrikosov et~al.}(1965)\citenamefont{Abrikosov, Gorkov,
  and Dzyaloshinskii}}]{Abrikosov65}
\bibinfo{author}{\bibfnamefont{A.}~\bibnamefont{Abrikosov}},
  \bibinfo{author}{\bibfnamefont{L.}~\bibnamefont{Gorkov}}, \bibnamefont{and}
  \bibinfo{author}{\bibfnamefont{I.}~\bibnamefont{Dzyaloshinskii}},
  \emph{\bibinfo{title}{Methods of Quantum Field Theory in Statistical
  Physics}} (\bibinfo{publisher}{Dover Publications}, \bibinfo{address}{New
  York}, \bibinfo{year}{1965}).

\bibitem[{\citenamefont{Kadanoff and Baym}(1962)}]{Kadanoff62}
\bibinfo{author}{\bibfnamefont{L.~P.} \bibnamefont{Kadanoff}} \bibnamefont{and}
  \bibinfo{author}{\bibfnamefont{G.}~\bibnamefont{Baym}},
  \emph{\bibinfo{title}{Quantum Statistical Mechanics}}
  (\bibinfo{publisher}{Benjamin}, \bibinfo{address}{New York},
  \bibinfo{year}{1962}).

\bibitem[{\citenamefont{Keldysh}(1965)}]{Keldysh65}
\bibinfo{author}{\bibfnamefont{L.}~\bibnamefont{Keldysh}},
  \bibinfo{journal}{Sov. Phys. JETP} \textbf{\bibinfo{volume}{20}},
  \bibinfo{pages}{1018} (\bibinfo{year}{1965}).

\bibitem[{\citenamefont{Schwinger}(1961)}]{Schwinger61}
\bibinfo{author}{\bibfnamefont{J.}~\bibnamefont{Schwinger}},
  \bibinfo{journal}{J. Math. Phys.} \textbf{\bibinfo{volume}{2}},
  \bibinfo{pages}{407} (\bibinfo{year}{1961}).

\bibitem[{\citenamefont{Ferguson et~al.}(2021)\citenamefont{Ferguson,
  Zilberberg, and Blatter}}]{Ferguson21}
\bibinfo{author}{\bibfnamefont{M.~S.} \bibnamefont{Ferguson}},
  \bibinfo{author}{\bibfnamefont{O.}~\bibnamefont{Zilberberg}},
  \bibnamefont{and} \bibinfo{author}{\bibfnamefont{G.}~\bibnamefont{Blatter}},
  \bibinfo{journal}{Phys. Rev. Research} \textbf{\bibinfo{volume}{3}},
  \bibinfo{pages}{023127} (\bibinfo{year}{2021}).

\bibitem[{\citenamefont{Sieberer}(2016)}]{Sieberer16}
\bibinfo{author}{\bibfnamefont{L.~M.} \bibnamefont{Sieberer}},
  \bibinfo{journal}{Rep. Prog. Phys.} \textbf{\bibinfo{volume}{79}},
  \bibinfo{pages}{096001} (\bibinfo{year}{2016}).

\bibitem[{\citenamefont{von Neumann}(1927)}]{vonNeumann27}
\bibinfo{author}{\bibfnamefont{J.}~\bibnamefont{von Neumann}},
  \bibinfo{journal}{G{\"o}ttinger Nachrichten} \textbf{\bibinfo{volume}{1}},
  \bibinfo{pages}{245} (\bibinfo{year}{1927}).

\bibitem[{\citenamefont{Landau}(1927)}]{Landau27}
\bibinfo{author}{\bibfnamefont{L.~D.} \bibnamefont{Landau}},
  \bibinfo{journal}{Phys. Z. Sowjetunion} \textbf{\bibinfo{volume}{45}},
  \bibinfo{pages}{430} (\bibinfo{year}{1927}).

\bibitem[{\citenamefont{Rammer and Smith}(1986)}]{Rammer86}
\bibinfo{author}{\bibfnamefont{J.}~\bibnamefont{Rammer}} \bibnamefont{and}
  \bibinfo{author}{\bibfnamefont{H.}~\bibnamefont{Smith}},
  \bibinfo{journal}{Rev. Mod. Phys.} \textbf{\bibinfo{volume}{58}},
  \bibinfo{pages}{323} (\bibinfo{year}{1986}).

\bibitem[{\citenamefont{Gaudin}(1960)}]{Gaudin60}
\bibinfo{author}{\bibfnamefont{M.}~\bibnamefont{Gaudin}},
  \bibinfo{journal}{Nuclear Physics} \textbf{\bibinfo{volume}{15}},
  \bibinfo{pages}{89} (\bibinfo{year}{1960}).

\bibitem[{\citenamefont{Zinn-Justin}(1989)}]{Zinn-Justin89}
\bibinfo{author}{\bibfnamefont{J.}~\bibnamefont{Zinn-Justin}},
  \emph{\bibinfo{title}{Quantum Field Theory and Critical Phenomena}}
  (\bibinfo{publisher}{Oxford University Press}, \bibinfo{address}{Oxford},
  \bibinfo{year}{1989}).

\bibitem[{\citenamefont{Kraus}(1971)}]{Kraus71}
\bibinfo{author}{\bibfnamefont{K.}~\bibnamefont{Kraus}}, \bibinfo{journal}{Ann.
  Phys.} \textbf{\bibinfo{volume}{64}}, \bibinfo{pages}{311}
  (\bibinfo{year}{1971}).

\bibitem[{\citenamefont{Gorini et~al.}(1976)\citenamefont{Gorini, Kossakowski,
  and Sudarshan}}]{Gorini76}
\bibinfo{author}{\bibfnamefont{V.}~\bibnamefont{Gorini}},
  \bibinfo{author}{\bibfnamefont{A.}~\bibnamefont{Kossakowski}},
  \bibnamefont{and} \bibinfo{author}{\bibfnamefont{E.~C.~G.}
  \bibnamefont{Sudarshan}}, \bibinfo{journal}{J. Math. Phys.}
  \textbf{\bibinfo{volume}{17}}, \bibinfo{pages}{821} (\bibinfo{year}{1976}).

\bibitem[{\citenamefont{Sudarshan}(1963)}]{Sudarshan63}
\bibinfo{author}{\bibfnamefont{E.~C.~G.} \bibnamefont{Sudarshan}},
  \bibinfo{journal}{Phys. Rev. Lett.} \textbf{\bibinfo{volume}{10}},
  \bibinfo{pages}{277} (\bibinfo{year}{1963}).

\bibitem[{\citenamefont{Chruscinski and Pascazio}(2017)}]{Chrus17}
\bibinfo{author}{\bibfnamefont{D.}~\bibnamefont{Chruscinski}} \bibnamefont{and}
  \bibinfo{author}{\bibfnamefont{S.}~\bibnamefont{Pascazio}},
  \bibinfo{journal}{Open Systems and Information Dynamics}
  \textbf{\bibinfo{volume}{24}}, \bibinfo{pages}{1740001}
  (\bibinfo{year}{2017}).

\bibitem[{\citenamefont{Whitney}(2008)}]{Whitney08}
\bibinfo{author}{\bibfnamefont{R.~S.} \bibnamefont{Whitney}},
  \bibinfo{journal}{J. Phys. A: Math. Theor.} \textbf{\bibinfo{volume}{41}},
  \bibinfo{pages}{175304} (\bibinfo{year}{2008}).

\bibitem[{\citenamefont{Cohen-Tannoudji
  et~al.}(1992)\citenamefont{Cohen-Tannoudji, Dupont-Roc, and
  Grynberg}}]{Cohen-Tannoudji92}
\bibinfo{author}{\bibfnamefont{C.}~\bibnamefont{Cohen-Tannoudji}},
  \bibinfo{author}{\bibfnamefont{J.}~\bibnamefont{Dupont-Roc}},
  \bibnamefont{and} \bibinfo{author}{\bibfnamefont{G.}~\bibnamefont{Grynberg}},
  \emph{\bibinfo{title}{Atom-Photon Interactions: Basic Processes and
  Applications}} (\bibinfo{publisher}{Wiley}, \bibinfo{address}{New York},
  \bibinfo{year}{1992}).

\bibitem[{\citenamefont{Diosi}(1990)}]{Diosi90}
\bibinfo{author}{\bibfnamefont{L.}~\bibnamefont{Diosi}},
  \bibinfo{journal}{Foundations of Physics} \textbf{\bibinfo{volume}{20}},
  \bibinfo{pages}{63} (\bibinfo{year}{1990}).

\bibitem[{\citenamefont{Diosi and Ferialdi}(1993)}]{Diosi93}
\bibinfo{author}{\bibfnamefont{L.}~\bibnamefont{Diosi}} \bibnamefont{and}
  \bibinfo{author}{\bibfnamefont{L.}~\bibnamefont{Ferialdi}},
  \bibinfo{journal}{Physica A} \textbf{\bibinfo{volume}{199}},
  \bibinfo{pages}{517} (\bibinfo{year}{1993}).

\bibitem[{\citenamefont{Hu et~al.}(1992)\citenamefont{Hu, Paz, and
  Zhang}}]{Hu92}
\bibinfo{author}{\bibfnamefont{B.~L.} \bibnamefont{Hu}},
  \bibinfo{author}{\bibfnamefont{J.~P.} \bibnamefont{Paz}}, \bibnamefont{and}
  \bibinfo{author}{\bibfnamefont{Y.}~\bibnamefont{Zhang}},
  \bibinfo{journal}{Phys. Rev. D} \textbf{\bibinfo{volume}{45}},
  \bibinfo{pages}{2843} (\bibinfo{year}{1992}).

\bibitem[{\citenamefont{Zwanzig}(1960)}]{Zwanzig60}
\bibinfo{author}{\bibfnamefont{R.}~\bibnamefont{Zwanzig}}, \bibinfo{journal}{J.
  Chem. Phys.} \textbf{\bibinfo{volume}{33}}, \bibinfo{pages}{1338}
  (\bibinfo{year}{1960}).

\bibitem[{\citenamefont{Zwanzig}(1964)}]{Zwanzig64}
\bibinfo{author}{\bibfnamefont{R.}~\bibnamefont{Zwanzig}},
  \bibinfo{journal}{Physica} \textbf{\bibinfo{volume}{30}},
  \bibinfo{pages}{1109} (\bibinfo{year}{1964}).

\bibitem[{\citenamefont{Butanas and Caballar}(2017)}]{Butanas17}
\bibinfo{author}{\bibfnamefont{B.~M.} \bibnamefont{Butanas}} \bibnamefont{and}
  \bibinfo{author}{\bibfnamefont{R.~C.~F.} \bibnamefont{Caballar}},
  \bibinfo{journal}{AIP Conference Proceedings 1871} p. \bibinfo{pages}{020006}
  (\bibinfo{year}{2017}).

\bibitem[{\citenamefont{Ivanov and Breuer}(2015)}]{Ivanov15}
\bibinfo{author}{\bibfnamefont{A.}~\bibnamefont{Ivanov}} \bibnamefont{and}
  \bibinfo{author}{\bibfnamefont{H.~P.} \bibnamefont{Breuer}},
  \bibinfo{journal}{Phys. Rev. A} \textbf{\bibinfo{volume}{92}},
  \bibinfo{pages}{032113} (\bibinfo{year}{2015}).

\bibitem[{\citenamefont{Smirne and Vacchini}(2010)}]{Smirne10}
\bibinfo{author}{\bibfnamefont{A.}~\bibnamefont{Smirne}} \bibnamefont{and}
  \bibinfo{author}{\bibfnamefont{B.}~\bibnamefont{Vacchini}},
  \bibinfo{journal}{Phys. Rev. A} \textbf{\bibinfo{volume}{82}},
  \bibinfo{pages}{022110} (\bibinfo{year}{2010}).

\bibitem[{\citenamefont{Breuer and Kappler}(2001)}]{Breuer01}
\bibinfo{author}{\bibfnamefont{H.~P.} \bibnamefont{Breuer}} \bibnamefont{and}
  \bibinfo{author}{\bibfnamefont{B.}~\bibnamefont{Kappler}},
  \bibinfo{journal}{Ann. Phys. (NY)} \textbf{\bibinfo{volume}{291}},
  \bibinfo{pages}{36} (\bibinfo{year}{2001}).

\bibitem[{\citenamefont{Chaturvedi and Shibata}(1979)}]{Chaturvedi79}
\bibinfo{author}{\bibfnamefont{S.}~\bibnamefont{Chaturvedi}} \bibnamefont{and}
  \bibinfo{author}{\bibfnamefont{F.}~\bibnamefont{Shibata}},
  \bibinfo{journal}{Z. Phys. B} \textbf{\bibinfo{volume}{35}},
  \bibinfo{pages}{297} (\bibinfo{year}{1979}).

\bibitem[{\citenamefont{Shibata et~al.}(1977)\citenamefont{Shibata, Takahashi,
  and Hashitsume}}]{Shibata77}
\bibinfo{author}{\bibfnamefont{F.}~\bibnamefont{Shibata}},
  \bibinfo{author}{\bibfnamefont{Y.}~\bibnamefont{Takahashi}},
  \bibnamefont{and}
  \bibinfo{author}{\bibfnamefont{N.}~\bibnamefont{Hashitsume}},
  \bibinfo{journal}{J. Stat. Phys.} \textbf{\bibinfo{volume}{17}},
  \bibinfo{pages}{171} (\bibinfo{year}{1977}).

\bibitem[{\citenamefont{Shibata and Arimitsu}(1980)}]{Shibata80}
\bibinfo{author}{\bibfnamefont{F.}~\bibnamefont{Shibata}} \bibnamefont{and}
  \bibinfo{author}{\bibfnamefont{T.}~\bibnamefont{Arimitsu}},
  \bibinfo{journal}{J. Phys. Soc. Jpn.} \textbf{\bibinfo{volume}{49}},
  \bibinfo{pages}{891} (\bibinfo{year}{1980}).

\bibitem[{\citenamefont{Xu et~al.}(2018)\citenamefont{Xu, Yan, and Liu}}]{Xu18}
\bibinfo{author}{\bibfnamefont{M.}~\bibnamefont{Xu}},
  \bibinfo{author}{\bibfnamefont{Y.}~\bibnamefont{Yan}}, \bibnamefont{and}
  \bibinfo{author}{\bibfnamefont{Y.}~\bibnamefont{Liu}}, \bibinfo{journal}{J.
  Chem. Phys} \textbf{\bibinfo{volume}{148}}, \bibinfo{pages}{164101}
  (\bibinfo{year}{2018}).

\bibitem[{\citenamefont{Venturi and Karniadakis}(2014)}]{Venturi14}
\bibinfo{author}{\bibfnamefont{D.}~\bibnamefont{Venturi}} \bibnamefont{and}
  \bibinfo{author}{\bibfnamefont{G.~E.} \bibnamefont{Karniadakis}},
  \bibinfo{journal}{Proc.R.Soc. A} \textbf{\bibinfo{volume}{470}},
  \bibinfo{pages}{20130754} (\bibinfo{year}{2014}).

\bibitem[{\citenamefont{te~Vrugt and Wittkowski}(2019)}]{teVrugt19}
\bibinfo{author}{\bibfnamefont{M.}~\bibnamefont{te~Vrugt}} \bibnamefont{and}
  \bibinfo{author}{\bibfnamefont{R.}~\bibnamefont{Wittkowski}},
  \bibinfo{journal}{Phys. Rev. E} \textbf{\bibinfo{volume}{99}},
  \bibinfo{pages}{062118} (\bibinfo{year}{2019}).

\bibitem[{\citenamefont{te~Vrugt and Wittkowski}(2020)}]{teVrugt20}
\bibinfo{author}{\bibfnamefont{M.}~\bibnamefont{te~Vrugt}} \bibnamefont{and}
  \bibinfo{author}{\bibfnamefont{R.}~\bibnamefont{Wittkowski}},
  \bibinfo{journal}{European Journal of Physics} \textbf{\bibinfo{volume}{41}},
  \bibinfo{pages}{045101} (\bibinfo{year}{2020}).

\bibitem[{\citenamefont{te~Vrugt}(2022)}]{teVrugt21}
\bibinfo{author}{\bibfnamefont{M.}~\bibnamefont{te~Vrugt}},
  \bibinfo{journal}{European Journal for Philosophy of Science}
  \textbf{\bibinfo{volume}{12}}, \bibinfo{pages}{41} (\bibinfo{year}{2022}).

\bibitem[{\citenamefont{Teretenkov}(2019)}]{Teretenkov19}
\bibinfo{author}{\bibfnamefont{A.~E.} \bibnamefont{Teretenkov}},
  \bibinfo{journal}{Lobachevskii Journal of Mathematics}
  \textbf{\bibinfo{volume}{40}}, \bibinfo{pages}{1587} (\bibinfo{year}{2019}).

\bibitem[{\citenamefont{Reimer et~al.}(2019)\citenamefont{Reimer, Wegewijs,
  Nestmann, and Pletyukhov}}]{Reimer19b}
\bibinfo{author}{\bibfnamefont{V.}~\bibnamefont{Reimer}},
  \bibinfo{author}{\bibfnamefont{M.~R.} \bibnamefont{Wegewijs}},
  \bibinfo{author}{\bibfnamefont{K.}~\bibnamefont{Nestmann}}, \bibnamefont{and}
  \bibinfo{author}{\bibfnamefont{M.}~\bibnamefont{Pletyukhov}},
  \bibinfo{journal}{J. Chem. Phys.} \textbf{\bibinfo{volume}{151}},
  \bibinfo{pages}{044101} (\bibinfo{year}{2019}).

\bibitem[{\citenamefont{Nestmann and Wegewijs}(2021)}]{Nestmann21a}
\bibinfo{author}{\bibfnamefont{K.}~\bibnamefont{Nestmann}} \bibnamefont{and}
  \bibinfo{author}{\bibfnamefont{M.~R.} \bibnamefont{Wegewijs}},
  \bibinfo{journal}{Phys. Rev. B} \textbf{\bibinfo{volume}{104}},
  \bibinfo{pages}{155407} (\bibinfo{year}{2021}).

\bibitem[{\citenamefont{Nestmann et~al.}(2021)\citenamefont{Nestmann, Bruch,
  and Wegewijs}}]{Nestmann21b}
\bibinfo{author}{\bibfnamefont{K.}~\bibnamefont{Nestmann}},
  \bibinfo{author}{\bibfnamefont{V.}~\bibnamefont{Bruch}}, \bibnamefont{and}
  \bibinfo{author}{\bibfnamefont{M.~R.} \bibnamefont{Wegewijs}},
  \bibinfo{journal}{Phys. Rev. X} \textbf{\bibinfo{volume}{11}},
  \bibinfo{pages}{021041} (\bibinfo{year}{2021}).

\bibitem[{\citenamefont{Ignatyuk and Morozov}(2022)}]{Ignatyuk22}
\bibinfo{author}{\bibfnamefont{V.~V.} \bibnamefont{Ignatyuk}} \bibnamefont{and}
  \bibinfo{author}{\bibfnamefont{V.~G.} \bibnamefont{Morozov}},
  \bibinfo{journal}{Condensed Matter Physics} \textbf{\bibinfo{volume}{25}},
  \bibinfo{pages}{13302} (\bibinfo{year}{2022}).

\bibitem[{\citenamefont{Reimer and Wegewijs}(2019)}]{Reimer19}
\bibinfo{author}{\bibfnamefont{V.}~\bibnamefont{Reimer}} \bibnamefont{and}
  \bibinfo{author}{\bibfnamefont{M.~R.} \bibnamefont{Wegewijs}},
  \bibinfo{journal}{SciPost Phys.} \textbf{\bibinfo{volume}{7}},
  \bibinfo{pages}{1} (\bibinfo{year}{2019}).

\bibitem[{\citenamefont{Maniscalco and Petruccione}(2006)}]{Maniscalco06}
\bibinfo{author}{\bibfnamefont{S.}~\bibnamefont{Maniscalco}} \bibnamefont{and}
  \bibinfo{author}{\bibfnamefont{F.}~\bibnamefont{Petruccione}},
  \bibinfo{journal}{Phys. Rev. A} \textbf{\bibinfo{volume}{73}},
  \bibinfo{pages}{012111} (\bibinfo{year}{2006}).

\bibitem[{\citenamefont{Yu and Eberly}(2006)}]{Yu06}
\bibinfo{author}{\bibfnamefont{T.}~\bibnamefont{Yu}} \bibnamefont{and}
  \bibinfo{author}{\bibfnamefont{J.~H.} \bibnamefont{Eberly}},
  \bibinfo{journal}{Phys. Rev. Lett.} \textbf{\bibinfo{volume}{97}},
  \bibinfo{pages}{140403} (\bibinfo{year}{2006}).

\bibitem[{\citenamefont{Schoeller}(2009)}]{Schoeller09}
\bibinfo{author}{\bibfnamefont{H.}~\bibnamefont{Schoeller}},
  \bibinfo{journal}{Eur. Phys. J. Special Topics}
  \textbf{\bibinfo{volume}{168}}, \bibinfo{pages}{179} (\bibinfo{year}{2009}).

\bibitem[{\citenamefont{Doll et~al.}(2008)\citenamefont{Doll, Zueco, Wubs,
  Kohler, and Haenggi}}]{Doll08}
\bibinfo{author}{\bibfnamefont{R.}~\bibnamefont{Doll}},
  \bibinfo{author}{\bibfnamefont{D.}~\bibnamefont{Zueco}},
  \bibinfo{author}{\bibfnamefont{M.}~\bibnamefont{Wubs}},
  \bibinfo{author}{\bibfnamefont{S.}~\bibnamefont{Kohler}}, \bibnamefont{and}
  \bibinfo{author}{\bibfnamefont{P.}~\bibnamefont{Haenggi}},
  \bibinfo{journal}{Chemical Physics} \textbf{\bibinfo{volume}{347}},
  \bibinfo{pages}{243} (\bibinfo{year}{2008}).

\bibitem[{\citenamefont{Chou et~al.}(1985)\citenamefont{Chou, Su, Hao, and
  Yu}}]{Chou85}
\bibinfo{author}{\bibfnamefont{K.}~\bibnamefont{Chou}},
  \bibinfo{author}{\bibfnamefont{Z.}~\bibnamefont{Su}},
  \bibinfo{author}{\bibfnamefont{B.}~\bibnamefont{Hao}}, \bibnamefont{and}
  \bibinfo{author}{\bibfnamefont{L.}~\bibnamefont{Yu}}, \bibinfo{journal}{Phys.
  Rep.} \textbf{\bibinfo{volume}{118}}, \bibinfo{pages}{1}
  (\bibinfo{year}{1985}).

\end{thebibliography}
\end{document}